\documentclass{lmcs} %%% last changed 2014-08-20
\pdfoutput=1

\usepackage{lastpage}
\lmcsdoi{18}{2}{10}
\lmcsheading{}{\pageref{LastPage}}{}{}%
{May~03,~2021}{May~18,~2022}{}


\usepackage[colorinlistoftodos,bordercolor=orange,backgroundcolor=orange!20,linecolor=orange,textsize=scriptsize]{todonotes}
\usepackage{hyperref}

\theoremstyle{plain} %\crefname{satz}{Satz}{S\"atze}

\usepackage[utf8]{inputenc}
\usepackage{amsmath,amsfonts,amssymb}
\usepackage{mathtools}
\usepackage{color}
\usepackage{tikz}
\usepackage{tikz-3dplot}

\usepackage{inconsolata}

\newcommand{\etc}{etc}
\newcommand{\ie}{i.e.}
\newcommand{\eg}{e.g.}
\newcommand{\cf}{cf.}
\newcommand{\R}{\mathbb{R}}
\newcommand{\N}{\mathbb{N}}
\newcommand{\Pcal}{\mathcal{P}}
\newcommand{\Fcal}{\mathcal{F}}
\newcommand{\Hcal}{\mathcal{H}}
\newcommand{\scalar}[2]{\langle #1, #2 \rangle}
\newcommand{\Coq}{\textsc{Coq}}
\newcommand{\CoqPolyhedra}{\textrm{Coq-Polyhedra}}
\newcommand{\MathComp}{Mathematical Components}
\newcommand{\MathCompShort}{MathComp}
\newcommand{\relevantfile}[1]{\href{https://github.com/Coq-Polyhedra/Coq-Polyhedra/tree/LMCS-21/theories/#1}{\texttt{#1}}}

\definecolor{dkblue}{rgb}{0,0.1,0.5}
\definecolor{lightblue}{rgb}{0,0.5,0.5}
\definecolor{dkgreen}{rgb}{0,0.4,0}
\definecolor{dk2green}{rgb}{0.4,0,0}
\definecolor{dkviolet}{HTML}{932191}
\definecolor{dkpink}{rgb}{1.2,0,1.6}
\definecolor{iden}{HTML}{0332FF}
\definecolor{comment}{HTML}{B12122}

\usepackage{listings}

\let\C=\lstinline

\usepackage{etoolbox}
\expandafter\patchcmd\csname \string\lstinline\endcsname{%
  \leavevmode
  \bgroup
}{%
  \leavevmode
  \ifmmode\hbox\fi
  \bgroup
}{}{%
  \typeout{Patching of \string\lstinline\space failed!}%
}

\usetikzlibrary{calc}
\usetikzlibrary{arrows}
\usetikzlibrary{positioning}
\usetikzlibrary{tikzmark}
\usetikzlibrary{decorations.markings}
\usetikzlibrary{intersections}

\newcommand{\ropen}[1]{[#1)} % chktex 9

\begin{document}

\title[Formalizing the Face Lattice of Polyhedra]{Formalizing the Face Lattice of Polyhedra\rsuper*}
\titlecomment{{\lsuper*}An extended abstract of the paper has appeared in the proceedings of the conference IJCAR 2020.}

\author[X.~Allamigeon]{Xavier Allamigeon\lmcsorcid{0000-0002-0258-8018}\rsuper{a}}	%required chktex 8
\address{Inria \and CMAP, CNRS, Ecole Polytechnique, Institut Polytechnique de Paris, 91128 Palaiseau Cedex, France}	%required
\email{xavier.allamigeon@inria.fr}  %optional
\thanks{The first author was partially supported by the ANR CAPPS project (ANR-17-CE40-0018), and by a public grant as part of the Investissement d'avenir project, reference ANR-11-LABX-0056-LMH, LabEx LMH, in a joint call with Gaspard Monge Program for optimization, operations research and their interactions with data sciences.}	%optional chktex 8

\author[R.~D.~Katz]{Ricardo D. Katz\lmcsorcid{0000-0002-6750-2692}\rsuper{b}}	%optional chktex 8
\address{CIFASIS--CONICET, Bv. 27 de Febrero 210 bis, 2000 Rosario, Argentina}	%optional chktex 8
\email{katz@cifasis-conicet.gov.ar}  %optional
%\thanks{thanks 2, optional.}	%optional

\author[P.-Y.~Strub]{Pierre-Yves Strub\lmcsorcid{0000-0002-8196-7875}\rsuper{c}}	%optional chktex 8
\address{LIX, CNRS, Ecole Polytechnique, Institut Polytechnique de Paris, 91128 Palaiseau Cedex, France}	%optional
\email{pierre-yves@strub.nu}  %optional
\thanks{The third author was partially supported by the ANR SCRYPT project (ANR-18-CE25-0014).}	%optional chktex 8

%% etc.

%% required for running head on odd and even pages, use suitable
%% abbreviations in case of long titles and many authors:

%%%%%%%%%%%%%%%%%%%%%%%%%%%%%%%%%%%%%%%%%%%%%%%%%%%%%%%%%%%%%%%%%%%%%%%%%%%

%% the abstract has to PRECEDE the command \maketitle:
%% be sure not to issue the \maketitle command twice!

\begin{abstract}
  \noindent Faces play a central role in the combinatorial and computational aspects of polyhedra. In this paper, we present the first formalization of faces of polyhedra in the proof assistant \Coq{}. This builds on the formalization of a library providing the basic constructions and operations over polyhedra, including projections, convex hulls and images under linear maps. Moreover, we design a special mechanism which automatically introduces an appropriate representation of a polyhedron or a face, depending on the context of the proof. We demonstrate the usability of this approach by establishing some of the most important combinatorial properties of faces, namely that they constitute a family of graded atomistic and coatomistic lattices closed under interval sublattices. We also prove a theorem due to Balinski on the $d$-connectedness of the adjacency graph of polytopes of dimension $d$.
\end{abstract}

\maketitle

%% start the paper here:
\section*{Introduction}\label{S:one}

A face of a polyhedron is defined as the set of points reaching the maximum (or minimum) of a linear function over the polyhedron.
Faces are ubiquitous in the theory of polyhedra, and especially in the complexity analysis of optimization algorithms. As an illustration, the simplex method, one of the most widely used algorithms for solving linear programming, finds an optimal solution by iterating over the \emph{graph} of the polyhedron, \ie, the adjacency graph of vertices and edges, which respectively constitute the $0$- and $1$-dimensional faces. The problem of finding a pivoting rule, \ie, a way to iterate over the graph, which ensures to reach an optimal vertex in a polynomial number of steps, is a central problem in computational optimization, related with Smale's ninth problem for the twenty-first century~\cite{Smale98}. Faces of polyhedra are also involved in the worst-case complexity analysis of other optimization methods, such as interior point methods; see~\cite{SIAGA2018,DTZ08}. This has motivated several mathematical problems on the
combinatorics of faces, which are of independent interest. For example, the question of finding a polynomial bound on the diameter of the graphs of polyhedra (in the dimension and the number of defining inequalities) is still unresolved, despite recent progress~\cite{BonifasDCG2014,Borgwardt2018,Santos2012}. We refer to~\cite{Jesus2019} for a recent account on the subject.

Other applications of polyhedra and their faces arise in formal verification, in which passing from a representation by inequalities to a representation as the convex hull of finitely many points and vice versa, is a critical computational step.
The correctness analysis of
the algorithms solving this problem extensively relies on the understanding of the mathematical structure of faces, in particular of vertices, edges and facets (\ie, $1$-co-dimensional faces).

In this paper, we formalize a significant part of the properties of faces in the proof assistant~\Coq{}. As usually happens in the formalization of mathematics, one of the key difficulties is to find the right representation for objects in the proof assistant. For polyhedra and their faces, the choice of the representation depends on the context. In more detail, every polyhedron admits infinitely many descriptions by linear inequality systems. In mathematics textbooks, proofs are carried out by choosing one (often arbitrary) inequality system for a polyhedron $\Pcal$, and then manipulating the faces of $\Pcal$ or other subsequent polyhedra through inequality systems which derive from the one chosen for $\Pcal$. Proving that these are valid inequality systems is usually trivial for the reader, but not for the proof assistant. We make use of the so-called \emph{canonical structures} of~\Coq{} in order to achieve this step automatically. This allows us to obtain proof scripts which only focus on the relevant mathematical content, and which are closer to what mathematicians write.

Thanks to this approach, we show that the faces of a polyhedron $\Pcal$ form a finite lattice, in which the order is the set inclusion, the bottom and top elements are respectively the empty set and $\Pcal$, and the meet operation is the set intersection. We establish that the face lattice is both atomistic and coatomistic, meaning that every element is the join (resp.\ the meet) of a finite set of atoms (resp.\ coatoms). Atoms and coatoms respectively correspond to minimal and maximal elements distinct from the top and bottom elements. Moreover, we prove that the face lattice is graded, \ie, every maximal chain has the same length. Finally, we show that the family of face lattices of polytopes (convex hulls of finitely many points) is closed under taking interval sublattices, \ie, any such sublattice of the face lattice of a polytope is isomorphic to the face lattice of another polytope. As a consequence of that, we prove that any interval sublattice of height two is isomorphic to a diamond.

Formalizing these results requires the introduction of several important and non-trivial notions. First of all, our work relies on the construction of a library manipulating polyhedra, which provides all the basic operations over them, including intersections, projections, convex hulls, as well as special classes of polyhedra such as affine subspaces. Dealing with faces also requires to formalize the dimension of a polyhedron, and its relation with the dimension of its affine hull, \ie, the smallest affine subspace containing it. Some classes of faces also retain a particular attention, such as vertices, edges and facets. For instance, we formalize the vertex figure, which is a geometric construction to manipulate the faces containing a fixed vertex. A concrete outcome of the formalization of these notions over faces is a formal proof of Balinski's Theorem, which states that the graph of an $n$-dimensional polytope remains connected even if $n-1$ vertices are removed from it.

Throughout this work, we have drawn much inspiration from the textbooks of Schrijver~\cite{Schrijver86} and Ziegler~\cite{ziegler} to guide us in our approach. The source code of our formalization is done within the \CoqPolyhedra{} project, and is available at {\small\url{https://github.com/Coq-Polyhedra/Coq-Polyhedra/tree/LMCS-21}}, in the directory \C$theories$. The source code described in this paper is tagged with LMCS-21 in the git repository. The repository contains a README.md file with installation instructions. We rely on the \MathComp{} library~\cite{MathComp} (abridged \MathCompShort{} thereafter) for basic data structures such as finite sets, ordered fields, and vector spaces. This work fits in a larger project which aims at formalizing a substantial part of the theory of polyhedra in a library which could be used both by computer scientists and mathematicians. That is why we pay a special attention to the design of the library as well as its integration into existing hierarchies of mathematical structures like the one developed by the \MathCompShort{} project.

This paper is organized as follows. In Section~\ref{sec:polyhedra}, we present how we define the basic operations and constructions over polyhedra, such as projections, images under linear maps, and convex hulls. Section~\ref{sec:representation} deals with the central problem of finding an appropriate representation of faces, and explains how this leads to a seamless formalization of important properties of faces. Section~\ref{sec:dim} shows how this approach applies to the formalization of the affine hull and the dimension of polyhedra as well. Sections~\ref{sec:face_lattice} and~\ref{sec:balinski} demonstrate the practical usability of our approach, by presenting the formalization of the face lattice and its main characteristics, and a formal proof of Balinski's Theorem, respectively. Finally, we discuss related work in Section~\ref{sec:related_work}. A link to the relevant source files is given beside section titles in order to help the reader finding the results in the source code of the formalization.

\section{Constructing a Library Manipulating Polyhedra}\label{sec:polyhedra}

\subsection[The Quotient Type of Polyhedra]{\texorpdfstring{\tikzmark[baseline]{m1}}{}The Quotient Type of Polyhedra}%
\begin{tikzpicture}[remember picture,overlay,font=\scriptsize]%
\node[left=1.1cm, anchor=base east,text width=2cm,align=right] at (pic cs:m1) {\relevantfile{hpolyhedron.v}
\relevantfile{polyhedron.v}};%
\end{tikzpicture}%
We recall that a \emph{(convex) polyhedron} of $\R^n$ is defined as the intersection of finitely many \emph{halfspaces} $\{x \in \R^n \colon \scalar{\alpha}{x} \geq \beta \}$, where $\alpha \in \R^n$, $\beta \in \R$, and $\scalar{\cdot}{\cdot}$ is the Euclidean scalar product, \ie, $\scalar{y}{z} \coloneqq \sum_{1 \leq i \leq n} y_i z_i$. Equivalently, a polyhedron can be represented as the solution set of a linear affine system $A x \geq b$, where $A \in \R^{m \times n}$ and $b \in \R^m$, in which case each inequality $A_i x \geq b_i$ corresponds to a halfspace.

Throughout the paper, we use the variable \C$n : nat$ to represent the dimension of the ambient space. Instead of dealing with polyhedra over the reals, we introduce a variable \C$R : realFieldType$ which represents an abstract ordered field with decidable ordering (this construction is provided by the \MathCompShort{} library).

As we mentioned earlier, the representation by inequalities (or halfspaces) of a convex polyhedron $\Pcal$ is not unique. The first step in our work is to introduce a quotient structure, in order to define the basic operations (membership of a point, inclusion, \etc) regardless of the exact representation of the polyhedron. This quotient structure is based on a concrete type denoted  \C$'hpoly[R]_n$ (or simply \C$'hpoly_n$, when \C$R$ is clear from the context). The prefix letter ``h'' is taken from the terminology \emph{H-polyhedron} or \emph{H-representation} which is used to refer to representations by halfspaces. The elements of \C$'hpoly_n$ are records consisting of a matrix $A \in \R^{m \times n}$ and a vector $b \in \R^m$ representing the system $A x \geq b$:\begin{lstlisting}
Record |*hpoly*| := HPoly { m : nat; A : 'M_(m,n); b : 'cV_m }.
\end{lstlisting}
Here, \C$'M[R]_(m,n)$ and \C$'cV[R]_m$ (or simply \C$'M_(m,n)$ and \C$'cV_m$) are types of \MathCompShort{} respectively representing $\C{m}\times \C{n}$-matrices and column vectors of size \C$m$ with entries of type \C$R$.
We equip the type \C$'hpoly_n$ with a membership predicate stating that, given \C$P : 'hpoly_n$ and \C$x : 'cV_n$, we have \C$x \in P$ if and only if \C$x$ satisfies the system of inequalities represented by \C$P$.

Instead of defining a complete set of primitives over H-polyhedra and then lift them to the quotient structure, we have chosen to isolate a minimal yet sufficient set of basic primitives. It encompasses the inclusion relation \C$|*poly_subset*|: 'hpoly_n -> 'hpoly_n -> bool$ which satisfies the following two statements:
\begin{lstlisting}
Lemma |*poly_subsetP*| (P Q : 'hpoly_n) :
  reflect (forall x, x \in P -> x \in Q) (poly_subset P Q).
Lemma |*poly_subsetPn*| (P Q : 'hpoly_n) :
  reflect (exists2 x, x \in P & x \notin Q) (~~ (poly_subset P Q)).
\end{lstlisting}
where \C$reflect$ stands for the logical equivalence between two properties, \C$~~ b$ for the Boolean negation of \C$b$ of type \C$bool$, and \C$exists2 x, p x & q x$ for the existence of an \C$x$ which satisfies both predicates \C$p$ and \C$q$. We highlight that, whenever possible, we define properties as Boolean predicates. One of its benefits is to enjoy traditional classical logic facilities, while still being in an intuitionistic setting. The relation \C$reflect$ brought by \MathCompShort{} allows to pass from one world to the other. For instance, from the Boolean predicate \C$poly_subset P Q$ to the \C$Prop$ predicate \C$forall x, x \in P -> x \in Q$ in \C$Lemma poly_subsetP$ above.

The definition of the primitive \C$poly_subset$, as well as the other Boolean predicates of \C$hpolyhedron.v$, exploits the formalization of the simplex method which has been done previously in~\cite{AllamigeonKatzJAR2018}. In short, it allows us to check that each inequality $\scalar{\alpha}{x} \geq \beta$ in the system defining \C$Q$ is valid over \C$P$ by minimizing the linear function $x \mapsto \scalar{\alpha}{x}$ over \C$P$, and checking that the optimal value is greater than or equal to $\beta$. Thanks to the \C$poly_subset$ relation, we can define the fact that two H-polyhedra represent the same set:
\begin{lstlisting}
Definition |*poly_equiv*| (P Q : 'hpoly_n) := (poly_subset P Q) && (poly_subset Q P).
Lemma |*poly_equivP*| (P Q : 'hpoly_n) : reflect (P =i Q) (poly_equiv P Q).
\end{lstlisting}
where \C$P =i Q$ stands for the extentional equality of the membership predicate associated with \C$P$ and \C$Q$.

The quotient structure is built following the approach of~\cite{CohenITP2013}. It introduces a quotient type, denoted here by \C$'poly[R]_n$ (or simply \C$'poly_n$). Its elements are referred to as \emph{polyhedra} and represent equivalence classes of H-polyhedra. In practice, each polyhedron is a record formed by a canonical representative of the class, and the proof that the representative is indeed the canonical one. We point out that the notion of canonical representative has no special mathematical meaning or structure.

We define the membership predicate of each \C$P : 'poly_n$ as the membership predicate of its canonical representative. As expected, equality between two polyhedra of \C$'poly_n$ and extensional equality of their membership predicates are equivalent properties:
\begin{lstlisting}
Lemma |*poly_eqP*| (P Q : 'poly_n) : (P = Q) <-> (P =i Q).
\end{lstlisting}

\subsection[Operations over Polyhedra]{\texorpdfstring{\tikzmark[baseline]{m2}}{}Operations over Polyhedra}\label{subsec:operations}%
\begin{tikzpicture}[remember picture,overlay,font=\scriptsize]
\node[left=1.1cm, anchor=base east,text width=2cm,align=right] at (pic cs:m2) {\relevantfile{polyhedron.v}};
\end{tikzpicture}%
We first lift a number of basic primitives from the type \C$'hpoly_n$ to the quotient type \C$'poly_n$, including the subset relation \C$P `<=` Q$ and the intersection operation \C$P `&` Q$. The associated properties are also lifted by using the fact that the membership predicate of any element of \C$'hpoly_n$ is extentionally equivalent to the membership predicate of its equivalence class in \C$'poly_n$.

Even though we now work on the quotient type, we still need a way to build polyhedra from sets of inequalities. While H-polyhedra rely on inequality constraints under the matrix form,
we choose now to be closer to the mathematical definition of polyhedra as the intersection of finitely many halfspaces. To this end, we introduce the type \C$lrel[R]_n$ (or simply \C$lrel_n$ when \C$R$ is clear from the context), which is isomorphic to the cartesian product \C$'cV_n * R$ of vectors of size \C$n$ and elements of \C$R$. This type is used to construct linear affine inequalities or equalities. It is equipped with a structure of left-\C$R$-module in order to build linear combinations of such relations. In more detail, if \C$e$ represents the pair $(\alpha, \beta) \in \R^n \times \R$, then the polyhedron \C$[hs e]$ corresponds to the halfspace $\scalar{\alpha}{x} \geq \beta$, so we have:
\begin{lstlisting}
Lemma |*in_hs*| (e : lrel_n) (x : 'cV_n) : x \in [hs e] <-> '[e.1,x] >= e.2.
\end{lstlisting}
Similarly, the element \C$e$ is used to build a hyperplane (which corresponds to the intersection \C$[hs e] `&` [hs (-e)]$) denoted \C$[hp e]$:
\begin{lstlisting}
Lemma |*in_hp*| (e : lrel_n) (x : 'cV_n) : x \in [hp e] <-> '[e.1,x] = e.2.
\end{lstlisting}
In the lemmas above, the terms \C$e.1 : 'cV_n$ and \C$e.2 : R$ respectively stand for the first and second component of the pair formed by \C$e$, while \C$'[.,.]$ stands for the scalar product between two vectors (see the file \C$inner_product.v$ of the \CoqPolyhedra{} project for its definition and properties).

We can now construct polyhedra defined by finite sets of inequalities. With this aim, we use the type \C${fset lrel_n}$ of finite sets of elements of type \C$lrel_n$. Then, given any \C$base : {fset lrel_n}$, the polyhedron denoted by \C$'P(base)$ is defined as follows:
\begin{lstlisting}
Definition |*poly_of_base*| (base : {fset lrel_n}) : 'poly_n :=
  \meet_(e : base) [hs (val e)].
Notation "''P' ( base )" := (poly_of_base base).
\end{lstlisting}
Thus, the term defining \C$'P(base)$ corresponds to the intersection of the halfspaces \C$[hs e]$ for \C$e \in base$. We then in particular introduce the empty polyhedron \C$[poly0]$ and the full polyhedron \C$[polyT]$, which are defined by the inequality $1 \leq 0$ and by no inequality respectively. As we shall see in Section~\ref{sec:representation}, the formalization of faces requires us to manipulate polyhedra defined by systems mixing inequalities and equalities. We denote such a polyhedron by \C$'P^=(base; I)$, where both \C$base$ and \C$I$ are of type \C${fset lrel_n}$. It represents the intersection of the polyhedron \C$'P(base)$ with the hyperplanes \C$[hp e]$ for \C$e \in I$.

The cornerstone of more advanced constructions is the primitive \C$proj0$, which, given a polyhedron \C$Q : 'poly_(n.+1)$, builds its projection on the last \C$n$ components. This is carried out by implementing Fourier--Motzkin elimination algorithm (see~\eg{}~\cite[Chapter~12]{Schrijver86}). In short, this algorithm starts from a system of linear inequalities defining \C$Q$, and constructs pairwise combinations of them in order to eliminate the first variable. The result is that the new system of linear inequalities is a valid representation of the projected polyhedron. This is stated as follows: % chktex 8
\begin{lstlisting}
Theorem |*proj0P*| (Q : 'poly_(n.+1)) (x : 'cV_n) :
  reflect (exists2 y : 'cV_(n.+1), x = row' 0 y & y \in Q) (x \in proj0 Q).
\end{lstlisting}
where \C$row' 0 y : 'cV_n$ is the projection of \C$y$ on the last \C$n$ components.

This projection primitive allows us to construct many more polyhedra. For example, we can build the image of a polyhedron \C$P : 'poly_n$ by the linear map represented by a matrix \C$A : 'M_(k,n)$. This is obtained by embedding \C$P$ in a polyhedron of \C$'poly_(n + k)$ over the variables \C$x : 'cV_n$ and \C$y : 'cV_k$, intersecting it with the equality constraints \C$y = A *m x$, where \C$*m$ stands for the matrix product, and finally projecting it on the last \C$k$ components. We denote this image by \C$map_poly A P$, and it indeed satisfies:
\begin{lstlisting}
Lemma |*in_map_polyP*| (A : 'M_(k,n)) (P : 'poly_n) (y : 'cV_k) :
  reflect (exists2 (x : 'cV_n), y = A *m x & x \in P) (y \in map_poly A P).
\end{lstlisting}

The construction of the convex hull of finitely many points immediately follows. Indeed, the convex hull of the finite set $\{v^1, \dots, v^p\} \subset \R^n$ can be defined as the image of the simplex $\Delta_p \coloneqq \{\mu \in (\R_{\geq 0})^p \colon \sum_{i = 1}^p \mu_i = 1\}$ by the linear map $\mu \mapsto \sum_{i = 1}^p \mu_i v^i$ (thus, in the construction above \C$P$ is defined as $\Delta_p$ and \C$A$ as the matrix whose columns are the vectors $v^1, \dots, v^p$). We denote the convex hull by \C$conv V$ where \C$V : {fset 'cV_n}$ represents a finite set of points, and we obtain (\cf~\C$Lemma |*in_convP*|$) that \C$x \in conv V$ if and only if \C$x$ is a convex combination of the points of \C$V$. We express the latter property as the equality \C$x = combine w$ where \C$w$ has type \C${convex 'cV[R]_n -> R}$ and satisfies \C$finsupp w `<=` V$. This respectively means that \C$w$ is a finite support function from \C$'cV[R]_n$ to \C$R$ such that its values \C$(w v)$ are nonnegative and sum \C$1$, and its support is a subset of \C$V$. Then, the point \C$combine w$ is defined as the barycenter \C$\sum_v (w v) *: v$ of the points \C$v$ in the support of \C$w$ (so \C$(w v) *: v$ stands for the scalar multiplication of \C$v$ by \C$(w v)$). We point out that the convex hull constructor yields some other elementary yet essential constructions, such as that of segments between two points (denoted \C$[segm x & y]$ for \C$x y : 'cV_n$).

Finally, we recover some important results of polyhedral theory that were proved in~\cite{AllamigeonKatzJAR2018} as a consequence of the termination and correctness of the simplex method. In more detail, we lift a version of Farkas Lemma expressed on the type \C$'hpoly_n$, and then obtain the Strong Duality Theorem and the complementary slackness conditions (which are conditions characterizing the optimality of solutions of linear programs). We also obtain some separation results, such as:
\begin{lstlisting}
Theorem |*separation*| (V : {fset 'cV_n}) (x : 'cV_n) :
  x \notin conv V -> exists2 e, x \notin [hs e] & (conv V `<=` [hs e]).
\end{lstlisting}
which states that if \C$x$ does not belong to the convex hull of \C$V$, we can find a separating halfspace, \ie, a halfspace which contains \C$conv V$ but not \C$x$. We refer to \C$Section |*Separation*|$ and \C$Section |*Duality*|$ for further details on these results.

\section[Representing Faces of Polyhedra]{Representing Faces of Polyhedra}\label{sec:representation}%
\subsection{\texorpdfstring{\tikzmark[baseline]{m3}}{}Equivalent Definitions of Faces}\label{subsec:face_def}%
\begin{figure}[t]
\begin{center}
\begin{tikzpicture}[convex/.style={draw=none,fill=lightgray,fill opacity=0.7},convexborder/.style={very thick,black!80},point/.style={blue!50},level_set/.style={blue!50, ultra thick, dotted},>=stealth',lbl/.style={font=\footnotesize}]
\begin{scope}[scale=.6]
\clip (-0.5,-0.5) rectangle (10.5,8.5);

\draw[help lines,gray!40] (-1,-1) grid (11,10);
\draw[gray!70,->] (-1,0) -- (10.5,0) node[above left] {$x_1$};
\draw[gray!70,->] (0,-1) -- (0,8.5) node[below right] {$x_2$};

\coordinate (v1) at (2,1);
\coordinate (v2) at (6,2);
\coordinate (v3) at (8,6);
\coordinate (v4) at (3,8);
\coordinate (v5) at (1,5);

\fill[convex] (v1) -- (v2) -- (v3) -- (v4) -- (v5) -- cycle;
\draw[convexborder] (v1) --
node[below=0.75ex,draw,very thin,circle,font=\tiny,inner sep=1pt] {$5$}
(v2) -- node[left,lbl] {$\{4\}$}
node[right=0.75ex,draw,very thin,circle,font=\tiny,inner sep=1pt] {$4$} (v3)
--
node[above=0.75ex,draw,very thin,circle,font=\tiny,inner sep=1pt] {$3$}
(v4) --
node[left=0.75ex,draw,very thin,circle,font=\tiny,inner sep=1pt] {$2$}
(v5) --
node[right=-2.5ex,draw,very thin,circle,font=\tiny,inner sep=1pt] {$1$}
cycle;

\filldraw[point] (v1) circle (4pt);
\filldraw[point] (v2) circle (4pt) node[above left=0ex and -.4ex,lbl] {$\{4,5\}$};
\filldraw[point] (v3) circle (4pt) node[below left=-.5ex and 0.5ex,lbl] {$\{2,4\}$};
\filldraw[point] (v4) circle (4pt);
\filldraw[point] (v5) circle (4pt);

\draw[very thick,->,blue!50] (6.5,1.5) -- (8,-0.2);
\draw[very thick,->,blue!50] (8.5,6.3) -- (10.5,7.3);
\draw[very thick,->,black!80] (7.5,3.5) -- (9.5,2.5);
\end{scope}

\begin{scope}[shift={(10,2)}]
\node[text width=6cm,font=\normalfont] {
\[
\begin{aligned}
2x_1 + x_2 & \geq 5 \\[.5ex]
5 x_1 - 2 x_2 & \geq -1 \\[.5ex]
-2 x_1 - 5x_2 & \geq -46 \\[.5ex]
- 2x_1 + x_2 & \geq -10 \\[.5ex]
-x_1 + 4x_2 & \geq 2
\end{aligned}
\]
};
\begin{scope}[shift={(2.5,1)}]
\node[draw,circle,font=\tiny,inner sep=1pt] at (0,.35) {$1$};
\node[draw,circle,font=\tiny,inner sep=1pt] at (0,-0.3) {$2$};
\node[draw,circle,font=\tiny,inner sep=1pt] at (0,-.95) {$3$};
\node[draw,circle,font=\tiny,inner sep=1pt] at (0,-1.6) {$4$};
\node[draw,circle,font=\tiny,inner sep=1pt] at (0,-2.25) {$5$};
\end{scope}
\end{scope}
\end{tikzpicture}
\end{center}
\caption{A polyhedron, defined by the inequalities on the right, and its faces. The vertices ($0$-dim.~faces) are represented by blue dots, while the edges ($1$-dim.~faces) are depicted in black. Arrows correspond to linear functions associated with some of the faces, in the sense of Definition~\ref{def:face}. We also indicate beside them the set $I$ of the defining inequalities turned into equalities, as in Theorem~\ref{th:face_characterization}.}\label{fig:face}
\end{figure}
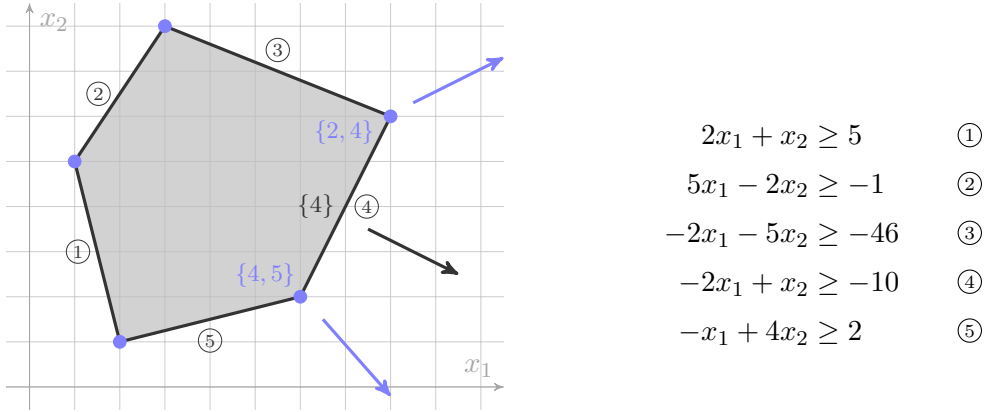

\begin{tikzpicture}[remember picture,overlay,font=\scriptsize]%
\node[above left=1cm and 1.1cm, anchor=base east,text width=2cm,align=right] at (pic cs:m3) {\relevantfile{poly\_base.v}};%
\end{tikzpicture}%
Faces are commonly defined as sets of optimal solutions of linear programs, \ie, problems consisting in minimizing a linear function over a polyhedron.
\begin{defi}\label{def:face}
A set $\Fcal$ is a \emph{face} of the polyhedron $\Pcal \subset \R^n$ if $\Fcal = \emptyset$ or there exists $c \in \R^n$ such that $\Fcal$ is the set of points of $\Pcal$ minimizing the linear function $x \mapsto \scalar{c}{x}$ over $\Pcal$.
\end{defi}
We note that $\Pcal$ is a face of itself (take $c = 0$). Figure~\ref{fig:face} provides an illustration of this definition.

In formal proving, the choice of the definition plays a major role on the ability to prove complex properties of the considered objects. A drawback of the previous definition is that it does not directly exhibit some of the most basic properties of faces: for instance, the fact that a face is itself a polyhedron, the fact that the intersection of two faces is a face, or the fact that a polyhedron has finitely many faces. In contrast, these properties are straightforward consequences of the following characterization of faces (we refer to Figure~\ref{fig:face} for an illustration of this characterization):
\begin{thm}\label{th:face_characterization}
Let $\Pcal = \{ x \in \R^n \colon A x \geq b \}$, where $A \in \R^{m \times n}$ and $b \in \R^m$. A set $\Fcal$ is a face of $\Pcal$ if and only if $\Fcal = \emptyset$ or there exists $I \subset \{1, \dots, m\}$ such that
\begin{equation}
\Fcal = \Pcal \cap \{ x \in \R^n \colon A_i x = b_i \enspace \text{for all} \enspace i \in I \} \, .\label{eq:face_characterization}
\end{equation}
\end{thm}
Nevertheless, Theorem~\ref{th:face_characterization} is expressed in terms of a certain H-representation of the polyhedron $\Pcal$, while the property of being a face is intrinsic to the set $\Pcal$. This raises the problem of exploiting the most convenient representation of $\Pcal$ to apply the characterization of Theorem~\ref{th:face_characterization}. We illustrate this on the proof of the following property, which is used systematically (or even implicitly) in almost every proof of statements on faces:
\begin{prop}\label{prop:face_of_face}
If $\Fcal$ is a face of $\Pcal$, then any face of $\Fcal$ is a face of $\Pcal$.
\end{prop}
Assume $\Pcal$ is represented by the inequality system $A x \geq b$, and take $I$ as in~\eqref{eq:face_characterization}. Let $\Fcal'$ be a nonempty face of $\Fcal$. We apply Theorem~\ref{th:face_characterization} with $\Fcal$ as $\Pcal$, by using  the following H-representation of $\Fcal$: $A x \geq b$ and $-A_i x \geq -b_i$ for $i \in I$. We get that $\Fcal' = \Fcal \cap \{ x \in \R^n \colon A_i x = b_i \; \text{for all} \; i \in I' \}$ for a certain set $I' \subset \{1, \dots, m\}$. We deduce that $\Fcal' = \Pcal \cap \{ x \in \R^n \colon A_i x = b_i \; \text{for all} \; i \in I \cup I' \}$, and conclude that $\Fcal'$ is a face of $\Pcal$ by applying Theorem~\ref{th:face_characterization}. While the choice of the H-representation of $\Pcal$ is irrelevant, we point out that the proof would not have been so immediate if we had initially chosen an arbitrary H-representation of $\Fcal$.

\subsection{Working within a Fixed Ambient H-Representation}\label{subsec:poly_base}

Theorem~\ref{th:face_characterization} leads us to the following strategy: when dealing with the faces of a polyhedron, and possibly with the faces of these faces, \etc, we first set an H-representation of the top polyhedron, and then manipulate the subsequent H-representations of faces in which some inequalities are strengthened into equalities, like in~\eqref{eq:face_characterization}.

The top H-representation will be referred to as the \emph{ambient representation}, and is formalized as a term \C$base$ of type \C${fset lrel_n}$ representing a finite set of inequalities. Then, we introduce the type \C${poly base}$, which corresponds to the subtype of \C$'poly_n$ whose inhabitants are the polyhedra \C$Q$ satisfying the following property:
\begin{lstlisting}
Definition |*has_base*| (base : {fset lrel_n}) (Q : 'poly_n) :=
  Q != [poly0] -> exists I : {fsubset base}, Q = 'P^=(base; I).
\end{lstlisting}
where \C${fsubset base}$ is the type of subsets of \C$base$. We recall that \C$'P^=(base; I)$ denotes the polyhedron defined by the inequalities in \C$base$, with the additional constraint that the inequalities in the subset \C$I$ are satisfied with equality. This means that \C${poly base}$ corresponds to the polyhedra defined by equalities or inequalities in \C$base$. The choice of the name \C$base$ is reminiscent of the terminology used in fiber bundles. Indeed, as we shall see in the next sections, several proofs will adopt the scheme of fixing a base locally, and then working on polyhedra of type \C${poly base}$. Following this analogy, the latter may be thought of as a fiber.

We can now present a first formalization of the set of faces relying on the subtype \C${poly base}$:
\begin{lstlisting}
Definition |*pb_face_set*| (P : {poly base}) := [set Q : {poly base} | Q `<=` P].
\end{lstlisting}
It defines the set of faces of \C$P : {poly base}$ as the set of elements of \C${poly base}$ contained in \C$P$. With this definition, some properties of faces come for free. For instance, the finiteness of the set of faces follows from the fact that there are only finitely many inhabitants of the type \C${fsubset base}$, and subsequently of \C${poly base}$. Another example is that Proposition~\ref{prop:face_of_face} straightforwardly derives from the transitivity of the subset relation \C$`<=`$.

Some other properties come at the price of proving that a polyhedron inhabits the type \C${poly base}$. As an example, if \C$P : {poly base}$ and \C$c : 'cV_n$, the term \C$argmin P c : 'poly_n$ represents the polyhedron formed by the set of points of \C$P$ minimizing the function \C$fun x => '[c,x]$. Since the inclusion \C$argmin P c `<=` P$ is immediate from the definition of \C$argmin P c$,
showing that \C$argmin P c$ is a face of \C$P$ essentially amounts to proving the following property:
\begin{lstlisting}
Lemma |*argmin_baseP*| (P : {poly base}) (c : 'cV_n) : has_base base (argmin P c).
\end{lstlisting}
The way to prove this statement is akin to the approach discussed in Section~\ref{subsec:face_def} for proving Proposition~\ref{prop:face_of_face}. Since \C$P$ is of type \C${poly base}$, it is of the form \C$P^=(base; I)$. In the case where \C$I$ is the empty set, the lemma above precisely corresponds to the ``only if'' part of Theorem~\ref{th:face_characterization}. We prove \C$Lemma |*argmin_baseP*|$ by first specializing to this case and exploiting the complementary slackness conditions. When \C$I$ is not empty, we simply reduce to the previous case by ``flattening'' the representation of \C$P$ to a set of inequalities only, \ie, writing it as \C$'P(base')$ where \C$base'$ is the union of \C$base$ with the opposite of the elements in \C$I$.

However, even once \C$Lemma |*argmin_baseP*|$ is proved, we cannot yet write a statement of the form \C$argmin P c \in pb_face_set P$ because \C$argmin P c$ has type \C$'poly_n$. In order to turn it into an element of the subtype \C${poly base}$, we need to explain in more detail how this type is defined. The type \C${poly base}$ is a short-hand notation for the following inductive type:
\begin{lstlisting}
Inductive |*poly_base*| (base : {fset lrel_n}) :=
  PolyBase { pval :> 'poly_n ; _ : has_base base pval }.
\end{lstlisting}
In other words, an element of type \C${poly base}$ is a record formed by an element \C$pval$ of type \C$'poly_n$ and a proof that the property \C$has_base base pval$ holds. While we could construct the element \C$PolyBase (argmin_baseP P c)$, we introduce a more general scheme to cast elements of type \C$'poly_n$ to \C${poly base}$ whenever possible. This scheme relies on \Coq{} canonical structures, which provide an automatic way to recover a term of record type from the head symbol. The association is declared as follows:
\begin{lstlisting}
Canonical |*argmin_base*| (P : {poly base}) (c : 'cV_n) :=
  PolyBase (argmin_baseP P c).
\end{lstlisting}
One restriction of \Coq{} is that canonical structures are resolved only when unifying types, and not arbitrary terms. This is why our primitive \C$poly_base_of$, which casts a \C$Q : 'poly_n$ to a \C${poly base}$, encapsulates the value \C$Q$ in a \emph{phantom type}, \ie, a type isomorphic to the unit type, but with a dependency on \C$Q$.
\begin{lstlisting}
Definition |*poly_base_of*| (Q : {poly base}) (_ : phantom 'poly_n Q) := Q.
Notation "Q %:poly_base" := (poly_base_of (Phantom _ Q)).
\end{lstlisting}
In consequence, writing \C$(argmin P c)%:poly_base$ triggers the unification algorithm between the term \C$argmin P c$ and a value of type \C${poly base}$, which is resolved using the \C$Canonical$ declared above. We finally end up with the following statement:
\begin{lstlisting}
Lemma |*argmin_pb_face_set*| (base : {fset lrel_n}) (P : {poly base}) (c : 'cV_n) :
  (argmin P c)%:poly_base \in pb_face_set P.
\end{lstlisting}
whose proof is trivial: it just amounts to proving the inclusion \C$argmin P c `<=` P$.

We declare other canonical structures over elementary constructions for which the property \C$has_base base _$ can be shown to be satisfied. This includes the intersection \C$P `&` Q$ of two polyhedra \C$P$ and \C$Q$ of type \C${poly base}$, the empty set \C$[poly0]$, or polyhedra of the form \C$'P(base)$ or \C$'P^=(base; .)$. This allows us to cast complex terms to the type \C${poly base}$, or, said differently, to prove automatically that they satisfy the property \C$has_base base _$. As an example, the term
\begin{lstlisting}
('P^=(base; I) `&` argmin 'P(base) c)%:poly_base : {poly base}
\end{lstlisting}
typechecks thanks to multiple resolutions of canonical structures on the aforementioned declarations, without requiring extra proof from the user. We refer to~\cite{Mahboubi2013} for the use of canonical structures in formal mathematics.

We point out that \C$Lemma |*argmin_pb_face_set*|$ corresponds to the ``only if'' part of the equivalence between the definition of faces brought by \C$pb_face_set$ and Definition~\ref{def:face} (\ie, the equivalence in Theorem~\ref{th:face_characterization}).
The ``if'' part can be stated as follows:
\begin{lstlisting}
Lemma |*pb_faceP*| (base : {fset lrel_n}) (P Q : {poly base}) :
  Q \in pb_face_set P -> Q != [poly0] -> exists c, Q = (argmin P c)%:poly_base.
\end{lstlisting}
When \C$Q$ is nonempty, we pick a set \C$I$ such that \C$Q = 'P^=(base, I)$, and build \C$c$ as the sum of the vectors \C$-e.1 : 'cV_n$ for \C$e \in I$. Then, the equality \C$Q = argmin P c$ follows from a routine verification of the complementary slackness conditions.

\subsection{Getting Free from Ambient Representations}\label{subsec:pseudo_quotient}

So far, we have worked with a fixed ambient representation \C$base : {fset lrel_n}$, and restricted the formalization of faces to polyhedra that can be expressed as terms of type \C${poly base}$. We now describe how to formalize the set of faces of any polyhedron of type \C$'poly_n$ as a finite set of polyhedra of the same type, without sacrificing the benefits brought by \C${poly base}$.

First, we exploit the observation that for each polyhedron \C$P : 'poly_n$, there exists \C$base : {fset lrel_n}$ and \C$P' : {poly base}$ such that \C$P = pval P'$ (recall that \C$pval$ also stands for the coercion from the type \C${poly base}$ to \C$'poly_n$). This can be proved by exploiting the definition of the quotient type \C$'poly_n$. Indeed, \C$P$ admits a representative \C$hrepr P : 'hpoly_n$ corresponding to a certain H-representation, from which we can build a term \C$base : {fset lrel_n}$ such that \C$P = pval 'P(base)%:poly_base$ holds.

Second, we introduce another quotient structure in order to deal with the fact that a polyhedron may correspond to several elements of type \C${poly base}$ for different  values of \C$base$. Our construction amounts to showing that \C$'poly_n$ is isomorphic to the quotient of the dependent sum type $\sum_{\C$base$} \C${poly base}$$ by the equivalence relation in which two polyhedra \C$Q1 : {poly base1}$ and \C$Q2 : {poly base2}$ are equivalent if \C$pval Q1 = pval Q2$. Given a polyhedron \C$P$ of type \C$'poly_n$, this construction provides us a canonical ambient representation denoted \C$\repr_base P : {fset lrel_n}$, and an associated canonical representative \C$\repr P$ of type \C${poly (\repr_base P)}$ satisfying \C$P = pval (\repr P)$.

We are now ready to define the set of faces of \C$P$ in full generality:
\begin{lstlisting}
Definition |*face_set*| (P : 'poly_n) :=
  [fset (pval F) | F in pb_face_set (\repr P)]%fset.
\end{lstlisting}
which corresponds to the image by the coercion \C$pval$ of the face set of \C$\repr P$ (here, \C$pval$ has type \C${poly (\repr_base P)} -> 'poly_n$). Of course, we need to check that this definition is independent of the choice of the representative of \C$P$ in this new quotient structure. This is stated as follows:
\begin{lstlisting}
Lemma |*face_set_morph*| (base : {fset lrel_n}) (P : {poly base}) :
  face_set P = [fset (pval F) | F in pb_face_set P]%fset.
\end{lstlisting}
The proof relies on the geometric properties of the elements of \C$pb_face_set$ established in Section~\ref{subsec:poly_base}. Indeed, they imply that, regardless of the choice of the ambient representation, the set \C$[fset (pval F) | F in pb_face_set P]$ always consists of the empty set \C$[poly0]$ and the polyhedra of the form \C$argmin P c$.

Now that this architecture is settled, we can prove some of the basic properties of faces. Most of the proofs make use of the following elimination principle:
\begin{lstlisting}
Lemma |*polybW*| (Pt : 'poly_n -> Prop) :
  (forall (base : {fset lrel_n}) (Q : {poly base}), Pt Q) ->
  	(forall P : 'poly_n, Pt P).
\end{lstlisting}
which states that, given a property to be proved for any polyhedron \C$P : 'poly_n$, it is sufficient to prove it for all \C$Q : {poly base}$ for any choice of \C$base$. In practice, the lemma above is used to introduce an ambient representation. Let us illustrate this on the proof of the fact that the intersection of two faces of \C$P$ is also a face of \C$P$:
\begin{lstlisting}
Lemma |*face_set_polyI*| (P Q1 Q2 : 'poly_n) :
  Q1 \in face_set P -> Q2 \in face_set P -> Q1 `&` Q2 \in face_set P.
Proof.
elim/polybW: P => base P.
case/face_setP => {}Q1 Q1_sub_P.
case/face_setP => {}Q2 Q2_sub_P.
by rewrite face_setE leIxl.
Qed.
\end{lstlisting}
The first line destructs \C$P$, introducing the ambient representation \C$base$ and an element of type \C${poly base}$ still named \C$P$. The second and third lines successively consume the assumptions that \C$Q1$ and \C$Q2$ are faces, introducing two elements of type \C${poly base}$ having the same name and respectively satisfying \C$Q1 `<=` P$ and \C$Q2 `<=` P$. Finally, the tactic \C$rewrite face_setE$ replaces \C$Q1 `&` Q2 \in face_set P$ by the goal \C$Q1 `&` Q2 `<=` P$, which is then proved by the order-theoretic \C$Lemma leIxl$ (over semilattices) from the inclusion \C$Q1 `<=` P$.

\section{From Faces to the Affine Hull and Dimension}\label{sec:dim}
\begin{tikzpicture}[remember picture,overlay,font=\scriptsize]%
\node[above left=.7cm and .3cm, anchor=base east,text width=2cm,align=right] at (pic cs:m5) {\relevantfile{affine.v}\\
\relevantfile{poly\_base.v}};
\end{tikzpicture}%
\tikzmark[baseline]{m5}We argue that the approach that we have introduced to represent faces of polyhedra also perfectly fits the formalization of the affine hull and dimension of polyhedra. Recall that the \emph{affine hull} of a polyhedron refers to the smallest (inclusionwise) affine subspace of $\R^n$ containing it, and the \emph{dimension} of the polyhedron is defined as the dimension of this subspace (\ie, the dimension of the underlying vector subspace).

\subsection{Active Inequalities, Affine Spaces and Affine Hull}

Given an ambient representation \C$base : {fset lrel_n}$ and a polyhedron \C$P : {poly base}$, we introduce the set \C${eq P}$ of \emph{active inequalities} of \C$P$ as follows:
\begin{lstlisting}
Definition |*active*| base (P : {poly base}) := \bigcup_(I | P `<=` 'P^=(base; I)) I.
Notation "'{eq'  P }" := (active P).
\end{lstlisting}
where \C$\bigcup$ stands for the union of finite sets.
The geometric interpretation of \C${eq P}$ is that it gathers all the linear relations \C$e \in base$ such that \C$P$ is contained in the corresponding hyperplane \C$[hp e]$, as stated in the following result:
\begin{lstlisting}
Lemma |*in_active*| base (P : {poly base}) e :
  e \in base -> (e \in {eq P} <-> P `<=` [hp e]%:PH).
\end{lstlisting}
Moreover, when \C$P$ is nonempty, \C${eq P}$ corresponds to the largest (inclusionwise) subset \C$I$ of \C$base$ such that \C$P = 'P^=(base; I)$.

It is a classic property of polyhedra that the affine hull of a nonempty polyhedron is the affine subspace defined by the intersection of the hyperplanes associated with its active inequalities. We take this property as a definition, and next lift it to the type \C$'poly_n$ using the quotient structure discussed in Section~\ref{subsec:pseudo_quotient}, just as we did with the set of faces:\begin{lstlisting}
Definition |*pb_hull*| base (P : {poly base}) : 'affine_n :=
  if P `>` [poly0] then [affine << {eq P} >>] else [affine0].
Definition |*hull*| (P : 'poly_n) := pb_hull (\repr P).
\end{lstlisting}
This definition makes use of a formalization of affine spaces we have built upon that of the vector spaces in the \MathCompShort{} library; see \relevantfile{affine.v}. In more detail, the type of affine spaces in dimension \C$n$ over the field \C$R$ is denoted \C$'affine[R]_n$ (or \C$'affine_n$ for short). Given a vector subspace \C$U : {vspace lrel_n}$ of \C$lrel_n$ (where the type constructor \C$vspace$ is provided by \MathCompShort{}), we define the affine space \C$[affine U] : 'affine_n$ as the set of points \C$x$ satisfying \C$'[e.1,x] = e.2$ for \C$e$ ranging over a basis of \C$U$. We then show that it coincides with the intersection of the hyperplanes represented by the elements of \C$U$:
\begin{lstlisting}
Lemma |*in_affineP*| (U : {vspace lrel_n}) x :
  reflect (forall e, e \in U -> '[e.1, x] = e.2) (x \in [affine U]).
\end{lstlisting}
so that the membership to \C$[affine U]$ is independent of the choice of the basis of \C$U$. In the definition of \C$pb_hull$ above, the term \C$<< {eq P} >>$ corresponds to the vector subspace of \C$lrel_n$ spanned by the finite set \C${eq P}$, while \C$[affine0]$ represents the empty affine space. An easy property that we can already obtain is that \C$pb_hull P$ (and subsequently \C$hull P$) contains \C$P$:
\begin{lstlisting}
Lemma |*pb_hull_subset*| base (P : {poly base}) : P `<=` (pb_hull P)%:PH.
\end{lstlisting}
(Here, \C$_%:PH$ is the function which casts any affine space to the polyhedron defined by the same equalities.) %XA: even though this notation is already used above, it is easier to explained once affine spaces have been discussed, so I prefer to keep it here.
This follows from the fact that \C$P = 'P^=(base; {eq P})$ whenever \C$P$ is not empty.

With this definition of the affine hull, we need to show that \C$hull P$ is independent of the choice of the ambient representation \C$base$, \ie:
\begin{lstlisting}
Lemma |*hullE*| base (P : {poly base}) : hull P = pb_hull P.
\end{lstlisting}
This is done by proving that \C$pb_hull P$ corresponds to the usual mathematical definition of the affine hull, \ie, it is included in any affine space \C$V$ that contains the polyhedron \C$P$:
\begin{lstlisting}
Lemma |*pb_hullP*| base (P : {poly base}) (V : 'affine_n) :
  (P `<=` V%:PH) <-> (pb_hull P `<=` V).
\end{lstlisting}
The fact that \C$Lemma pb_hullP$ implies \C$Lemma hullE$ can be established by using the inclusions \C$P `<=` (hull P)%:PH$ and \C$P `<=` (pb_hull P)%:PH$. Indeed, recall that \C$hull P$ is defined as \C$pb_hull (\repr P)$, and that \C$P$ and \C$\repr P$ represent the same polyhedron, \ie, \C$pval P = pval (\repr P)$. Thus, \C$Lemma pb_hullP$ applied with \C$P := P$ and \C$P := \repr P$ respectively yields the inclusions \C$pb_hull P `<=` hull P$ and \C$hull P `<=` pb_hull P$.

We comment on the proof of \C$Lemma pb_hullP$, since it makes use of the construction of a special point of \C$P$ of independent interest, which can be interpreted as being in its relative interior (\ie, the interior of \C$P$ w.r.t.~the topology induced on its affine hull). In what follows, we suppose that \C$P : {poly base}$ is nonempty, since the statement \C$Lemma pb_hullP$ is trivial otherwise. In this setting, for each \C$e$ in \C$base `\` {eq P}$ (the complement set of \C${eq P}$ in \C$base$) we build a point \C$y \in P$ that satisfies \C$'[e.1,y] > e.2$ by solving the linear program that maximizes the linear function \C$x => '[e.1,x]$ over \C$P$. Such a point must exist since otherwise we would have \C$'[e.1, x] = e.2$ for all \C$x \in P$, \ie, \C$P `<=` [hp e]$, which would contradict the assumption that \C$e \in base `\` {eq P}$ (\C$Lemma in_active$). Taking the isobarycenter of these points provides the point \C$relint_pt P$ that belongs to \C$P$ and satisfies the following property:
\begin{lstlisting}
Lemma |*relint_ptP*| base (P : {poly base}) e :
  e \in base `\` {eq P} -> relint_pt P \notin [hp e].
\end{lstlisting}
In other words, \C$relint_pt P$ satisfies all inactive inequalities of \C$P$ in a strict way. Now, let us consider an affine space \C$V$ such that \C$P `<=` V%:PH$. As \C$P$ is a nonempty set included in both \C$V$ and \C$pb_hull P$ (by \C$Lemma pb_hull_subset$), the inclusion \C$pb_hull P `<=` V$ reduces to the inclusion of the  direction spaces \C$dir (pb_hull P)$ and \C$dir V$.\footnote{We recall that the \emph{direction space} of an affine space $V$ corresponds to the vector space spanned by the elements $x - y$ for $x, y$ ranging over $V$.} The latter inclusion can be proved as follows. It is easy to see that any vector \C$d \in dir (pb_hull P)$ satisfies \C$'[e.1,d] = 0$ for all \C$e \in << {eq P} >>$.
For any sufficiently small scalar \C$t != 0$, we can show that the point \C$y := relint_pt P + t *: d$ still belongs to \C$P = 'P^=(base; {eq P})$ (recall that \C$t *: d$ stands for the scalar multiplication of the vector \C$d$ by \C$t$). Indeed, for \C$t$ sufficiently small we have \C$'[e.1, relint_pt P + t *: d] > e.2$ if \C$e \in base `\` {eq P}$ by \C$Lemma relint_ptP$, while \C$'[e.1, relint_pt P + t *: d] = '[e.1, relint_pt P] = e.2$ if \C$e \in {eq P}$. We finally deduce from the inclusion \C$P `<=` V%:PH$ that \C$y - relint_pt P = t *: d$ belongs to the vector space \C$dir V$, which entails that \C$d \in dir V$ as \C$t != 0$. This completes the proof of the implication \C$->$ in \C$Lemma pb_hullP$. The converse implication is a trivial consequence of the transitivity of the subset relation and \C$Lemma pb_hull_subset$.

Of course, we can also lift the statement \C$Lemma pb_hullP$ to the type \C$'poly_n$, which leads to the usual characteristic property of affine hulls:
\begin{lstlisting}
Lemma |*hullP*| (P : 'poly_n) (V : 'affine_n) :
  (P `<=` V%:PH) <-> (hull P `<=` V).
\end{lstlisting}

\subsection{The Dimension of Polyhedra}

The definition of the dimension of a polyhedron straightforwardly follows from that of its affine hull:
\begin{lstlisting}
Notation "\pdim P" := (adim (hull P)).
\end{lstlisting}
where \C$adim V$ stands for the dimension of the affine space \C$V : 'affine_n$. When \C$V$ is nonempty, its dimension is usually defined as the dimension of its direction space \C$dir V$, and it is a common convention to set the dimension of the empty set to $-1$. However, in our case, we want the dimension of affine spaces (and subsequently of polyhedra) to range over the type \C$nat$ of natural numbers. Therefore, we have to set the dimension of the empty affine space \C$[affine0]$ to $0$, and shift by one the dimension of nonempty affine spaces. This leads to the following definition:
\begin{lstlisting}
Definition |*adim*| V :=
  if V == [affine0] then 0 else (\dim (dir V)).+1.
\end{lstlisting}

We compute the dimension for important classes of polyhedra. To this end, it is convenient to use the constructor \C$[affine W & O] : 'affine_n$ which provides the affine space with origin \C$O$ and parallel to the vector subspace \C$W$, \ie, the set of points of the form \C$O + w$ for \C$w \in W$. (The direction space of such an affine space is simply given by \C$W$.)

As an example, we show that the affine hull of the segment \C$[segm x & y]$ is given by the line passing through \C$x$ and directed by the vector \C$y-x$:
\begin{lstlisting}
Lemma |*hull_line*| (x y : 'cV_n) :
  hull [segm x & y] = [affine <[y-x]> & x].
\end{lstlisting}
(where \C$<[y-x]>$ stands for the vector subspace spanned by \C$y-x$). This in turns implies that such a segment has dimension \C$2$ or \C$1$, depending on whether \C$x != y$ (remember the shift by one of our formalization):
\begin{lstlisting}
Lemma |*dim_segm*| (x y : 'cV_n) : \pdim [segm x & y] = (x != y).+1.
\end{lstlisting}
Conversely, we show that any compact polyhedron of dimension $2$ is a segment of two distinct points:
\begin{lstlisting}
Lemma |*dim2P*| (P : 'poly_n) :
  compact P -> \pdim P = 2 -> exists x, exists2 y, P = [segm x & y] & x != y.
\end{lstlisting}
Here, \C$compact P$ is simply defined as the fact that \C$P$ is a bounded set, as polyhedra are topologically closed. The proof of \C$Lemma dim2P$ is usually omitted in mathematics textbooks because it is geometrically obvious. However, what is trivial for mathematicians is often not so trivial for proof assistants, and here, we explicitly build the points \C$x$ and \C$y$ by minimizing and maximizing the linear function \C$fun z => '[d,z]$ over \C$P$, where \C$d$ is any nonzero element of the one-dimensional vector subspace \C$dir (hull P)$.

In a similar way, we prove that polyhedra reduced to a single point are precisely the ones having dimension $1$ (\C$Lemma |*dim_pt*|$ and \C$Lemma |*dim1P*|$), that proper hyperplanes have codimension $1$ (\C$Lemma |*dim_hp*|$), \etc. We refer to \C$Section |*Dimension*|$ for a detailed account of these results.

\subsection{Facets of Polyhedra}

We close this section by discussing the formalization of facets of polyhedra and their combinatorial characterization in terms of active inequalities. We recall that a \emph{facet} of a nonempty polyhedron $\Pcal$ is a face of $\Pcal$ of dimension $\dim \Pcal - 1$. % chktex 8

A classical result states that when $\Pcal$ is given by a \emph{non-redundant} system of inequalities $A x \geq b$ (\ie, the H-representation is minimal inclusionwise), the facets are precisely the polyhedra of the form $\Pcal \cap \{ x \in \R^n \colon A_i x = b_i\}$ for any $i$ such that $\Pcal \not \subset \{ x \in \R^n \colon A_i x = b_i\}$. The formalization of this statement first goes through the construction of non-redundant bases for any polyhedron, and the proof of the following elimination principle:
\begin{lstlisting}
Lemma |*non_redundant_baseW*| (Pt : 'poly_n -> Prop) :
  (forall base, non_redundant base -> Pt 'P(base)%:poly_base) ->
  	(forall P : 'poly_n, Pt P).
\end{lstlisting}
This allows to specialize \C$P$ to a polyhedron of the form \C$'P(base)$ where \C$base$ is a minimal set of inequalities defining \C$P$. Using the techniques of Section~\ref{sec:representation}, we switch to a proof environment dealing with polyhedra in \C${poly base}$, and establish that the facets of \C$P$ are precisely the polyhedra of the form \C$'P^=(base; [fset e])$ for \C$e \notin {eq P}$ (recall that \C$[fset e]$ is the singleton consisting of the element \C$e$). This is done by showing that the set of active inequalities of such a polyhedron is precisely the union \C${eq P} `|` [fset e]$ of \C${eq P}$ and \C$[fset e]$:
\begin{lstlisting}
Lemma |*activeU1*| e :
  e \in base -> {eq 'P^=(base; [fset e])%:poly_base } = {eq P} `|` [fset e].
\end{lstlisting}
so that the affine hull of \C$'P^=(base; [fset e])$ reduces to the intersection of \C$hull P$ with the hyperplane \C$[hp e]$. When \C$e \notin {eq P}$, this affine space has dimension \C$(dim P)-1$, which leads to the expected statement:
\begin{lstlisting}
Lemma |*dim_facet*| e :
  e \in base `\` {eq P} -> \pdim P = (\pdim 'P^=(base; [fset e])%:poly_base).+1%N.
\end{lstlisting}
We refer to \C$Lemma |*facetP*|$ for the converse statement, in which we show that $1$-co-dimensional faces of \C$P$ are exactly the polyhedra of the form \C$'P^=(base; [fset e])$ for \C$e$ ranging in \C$base `\` {eq P}$, whose proof relies on analogous arguments.

\section[The face lattice]{The Face Lattice}\label{sec:face_lattice}
\tikzmark[baseline]{m6}In this section, we illustrate how the framework that we have introduced in Sections~\ref{sec:representation} and~\ref{sec:dim} serves as a foundation for formalizing the structural properties of faces. We refer to Figure~\ref{fig:face_lattice} for an example of the properties presented below.
\begin{tikzpicture}[remember picture,overlay,font=\scriptsize]%
\node[above left=0.7cm and 0.3cm, anchor=base east,text width=2cm,align=right] at (pic cs:m6) {\relevantfile{poly\_base.v}};%
\end{tikzpicture}%

At the core of the formalization lies the theory of ordered structures such as partial orders, semilattices and lattices. Some of these structures have been very recently introduced in the \MathCompShort{} library --~for instance, the non-distributive lattice structure has been introduced in early 2020. However, as we shall see in this section, the formalization of the face lattice requires the implementation of additional objects, such as graded lattices, interval sublattices, and lattice homomorphisms. This development is gathered in the module \C$xorder.v$ of the \CoqPolyhedra{} project.

The first property that we can immediately formalize following the results of Section~\ref{sec:representation} is the finite lattice structure over the set \C$face_set P$ for \C$P : 'poly_n$. The partial order is given by the polyhedron inclusion \C$`<=`$, the meet operator is the intersection \C$`&`$ (as a consequence of \C$Lemma |*face_set_polyI*|$, see Section~\ref{subsec:pseudo_quotient}), while the bottom and top elements are respectively \C$[poly0]$ and \C$P$. As a finite lattice, the join operator \C$Q `|` Q'$ can be built as the meet of all the faces of \C$P$ containing both \C$Q$ and \C$Q'$. We use a dedicated construction to build such a lattice from a finite partially ordered set endowed with a meet operator and bottom/top elements (see \C$Section MeetBTFinMixin$ in \C$xorder.v$).

\begin{figure}[t]
\begin{center}
\begin{tikzpicture}[scale=1.2,general/.style={fill=lightgray,draw=black,thin,fill opacity=.4},facet/.style={fill=blue!50},edge/.style={thick,fill opacity=0.4,line cap=round},vertex/.style={orange},back/.style={black!80,draw opacity=0.5,dotted},front/.style={black!80},lbl/.style={font=\footnotesize}]

\begin{scope}[scale=1]
\coordinate (v000) at (0,0,0);
\path (v000) node[above right,lbl] {$2$};
\coordinate (v001) at (0,0,2);
\path (v001) node[below left=0.2ex and 0ex,lbl] {$5$};
\coordinate (v010) at (0,2,0);
\path (v010) node[above right,lbl] {$3$};
\coordinate (v011) at (0,2,2);
\path (v011) node[below right=0.1ex and 0ex,lbl] {$4$};
\coordinate (v) at (3,1,1);
\path (v) node[right,lbl] {$1$};
\coordinate (w) at (-2,1,0);
\path (w) node[above left,lbl] {$6$};
\coordinate (w') at (-2,1,1);
\path (w') node[below left,lbl] {$7$};

\draw[edge,back] (v000) -- (v001);
\draw[edge,front] (v001) -- (v011);
\draw[edge,front] (v011) -- (v010);
\draw[edge,back] (v010) -- (v000);
\draw[edge,back] (v) -- (v000);
\draw[edge,front] (v) -- (v001);
\draw[edge,front] (v) -- (v011);
\draw[edge,front] (v) -- (v010);
\draw[edge,back] (w) -- (v000);
\draw[edge,front] (w') -- (v001);
\draw[edge,front] (w') -- (v011);
\draw[edge,front] (w) -- (v010);
\draw[edge,front] (w) -- (w');
\end{scope}

\begin{scope}[shift={(5,-3)},scale=0.65]

\coordinate (bot) at (0,-1.2);

\coordinate (v1) at (-3,0);
\coordinate (v2) at (-2,0);
\coordinate (v3) at (-1,0);
\coordinate (v4) at (0,0);
\coordinate (v5) at (1,0);
\coordinate (v6) at (2,0);
\coordinate (v7) at (3,0);

\coordinate (e12) at (-6,1);
\coordinate (e13) at (-5,1);
\coordinate (e14) at (-4,1);
\coordinate (e15) at (-3,1);
\coordinate (e23) at (-2,1);
\coordinate (e34) at (-1,1);
\coordinate (e45) at (0,1);
\coordinate (e52) at (1,1);
\coordinate (e26) at (2,1);
\coordinate (e36) at (3,1);
\coordinate (e47) at (4,1);
\coordinate (e57) at (5,1);
\coordinate (e67) at (6,1);

\coordinate (f123) at (-3.5,2);
\coordinate (f134) at (-2.5,2);
\coordinate (f145) at (-1.5,2);
\coordinate (f125) at (-0.5,2);
\coordinate (f236) at (0.5,2);
\coordinate (f3467) at (1.5,2);
\coordinate (f2567) at (2.5,2);
\coordinate (f457) at (3.5,2);

\coordinate (top) at (0,3.2);

\draw[black!80] (bot) -- (v1) (bot) -- (v2) (bot) -- (v3) (bot) -- (v4) (bot) -- (v5) (bot) -- (v6) (bot) -- (v7);

\draw[black!80] (v1) -- (e12) -- (v2) (v1) -- (e13) -- (v3) (v1) -- (e14) -- (v4) (v1) -- (e15) -- (v5);
\draw[black!80] (v2) -- (e23) -- (v3);
\draw[blue!50,thick] (v3) -- (e34);
\draw[black!80] (e34) -- (v4) -- (e45) -- (v5) -- (e52) -- (v2);
\draw[black!80] (v2) -- (e26) -- (v6);
\draw[blue!50,thick] (v3) -- (e36);
\draw[black!80] (e36) -- (v6) (v4) -- (e47) -- (v7) (v5) -- (e57) -- (v7) (v6) -- (e67) -- (v7);

\draw[black!80] (f123) -- (e12) (f123) -- (e13) (f123) -- (e23);
\draw[black!80] (f134) -- (e13) (f134) -- (e14) (f134) -- (e34);
\draw[black!80] (f145) -- (e14) (f145) -- (e15) (f145) -- (e45);
\draw[black!80] (f125) -- (e12) (f125) -- (e15) (f125) -- (e52);
\draw[black!80] (f236) -- (e23) (f236) -- (e36) (f236) -- (e26);
\draw[blue!50,thick] (f3467) -- (e34);
\draw[black!80] (f3467) -- (e47) (f3467) -- (e67);
\draw[blue!50,thick] (f3467) -- (e36);
\draw[black!80] (f2567) -- (e52) (f2567) -- (e57) (f2567) -- (e67) (f2567) -- (e26);
\draw[black!80] (f457) -- (e45) (f457) -- (e57) (f457) -- (e47);

\draw[black!80] (top) -- (f123) (top) -- (f134) (top) -- (f145) (top) -- (f125) (top) -- (f236) (top) -- (f3467) (top) -- (f2567) (top) -- (f457);

\node[lbl,anchor=east] at (-6.7,0) {vertices};
\draw[dotted] (-6.5,0) -- (6.5,0);
\node[lbl,anchor=east] at (-6.7,1) {edges};
\draw[dotted] (-6.5,1) -- (6.5,1);
\node[lbl,anchor=east] at (-6.7,2) {facets};
\draw[dotted] (-6.5,2) -- (6.5,2);

\filldraw (bot) circle (2pt);
\filldraw (v1) circle (2pt);
\path (v1) node[below,lbl] {$1$};
\filldraw (v2) circle (2pt);
\path (v2) node[below,lbl] {$2$};
\filldraw[blue!50] (v3) circle (2pt);
\path (v3) node[below,lbl] {$3$};
\filldraw (v4) circle (2pt);
\path (v4) node[below,lbl] {$4$};
\filldraw (v5) circle (2pt);
\path (v5) node[below,lbl] {$5$};
\filldraw (v6) circle (2pt);
\path (v6) node[below,lbl] {$6$};
\filldraw (v7) circle (2pt);
\path (v7) node[below,lbl] {$7$};

\filldraw (e12) circle (2pt);
\filldraw (e13) circle (2pt);
\filldraw (e14) circle (2pt);
\filldraw (e15) circle (2pt);
\filldraw (e23) circle (2pt);
\filldraw[blue!50] (e34) circle (2pt);
\filldraw (e45) circle (2pt);
\filldraw (e52) circle (2pt);
\filldraw (e26) circle (2pt);
\filldraw[blue!50] (e36) circle (2pt);
\filldraw (e47) circle (2pt);
\filldraw (e57) circle (2pt);
\filldraw (e67) circle (2pt);

\filldraw (f123) circle (2pt);
\filldraw (f134) circle (2pt);
\filldraw (f145) circle (2pt);
\filldraw (f125) circle (2pt);
\filldraw (f236) circle (2pt);
\filldraw[blue!50] (f3467) circle (2pt);
\filldraw (f2567) circle (2pt);
\filldraw (f457) circle (2pt);
\filldraw (top) circle (2pt);
\end{scope}

\end{tikzpicture}
\end{center}
\caption{A three-dimensional polytope (left) and the Hasse diagram of its face lattice (right). A interval of height $2$ is depicted in blue.}\label{fig:face_lattice}
\end{figure}
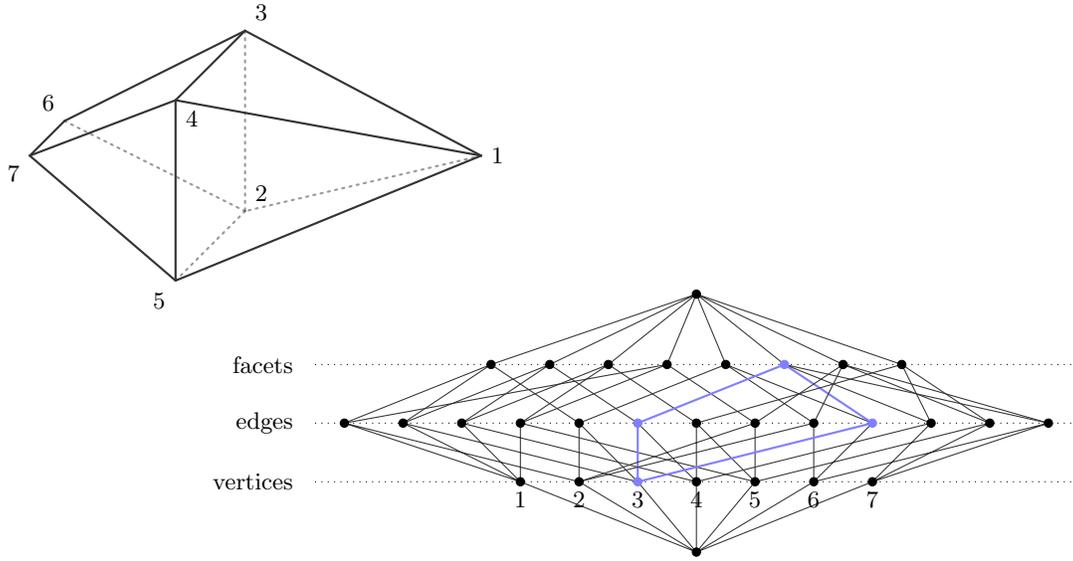

\subsection{Facets and Gradedness}\label{sec:gradedness}

Recall that a lattice $(L, \prec)$ is said to be \emph{graded} if there exists a rank function $\Phi \colon L \to \N$ such that:
\begin{enumerate}
\item\label{item:graded1} $\Phi(u) < \Phi(v)$ whenever $u \prec v$,
\item\label{item:graded2} $u \preceq v$ and $\Phi(u)+1 < \Phi(v)$ implies that there exists $w \in L$ satisfying $u \prec w \prec v$.
\end{enumerate}
Equivalently, this is a lattice in which all maximal chains have the same length.

In the case of the face lattice, the rank function can be defined as the dimension of the face. Property~\eqref{item:graded1} is proved as follows. If \C$Q$ and \C$Q'$ are both faces of \C$P$ and \C$Q `<` Q'$, we have \C$dim Q <= dim Q'$. Indeed, as a consequence of \C$Lemma hullP$, we have \C$hull Q `<=` hull Q'$, so that  the dimension is monotone (\C$Lemma |*dimS*|$). Moreover, if we assume \C$dim Q = dim Q'$, then we can prove that \C$hull Q$ is equal to \C$hull Q'$ (as affine subspaces of the same dimension). We conclude that \C$Q = Q'$ by using the following lemma:
\begin{lstlisting}
Lemma |*face_hullI*| (P F : 'poly_n) :
  F \in face_set P -> F = P `&` (hull F)%:PH.
\end{lstlisting}
applied to the faces \C$Q$ and \C$Q'$. Property~\eqref{item:graded2} is stated as follows:
\begin{lstlisting}
Lemma |*graded*| (P Q Q' : 'poly_n) :
  Q \in face_set P -> Q' \in face_set P ->
    Q `<=` Q' -> (\pdim Q).+1 < \pdim Q' ->
      exists2 F, F \in face_set P & Q `<` F `<` Q'.
\end{lstlisting}
and its proof relies on the formalization of facets. We first consider the case where \C$Q' = P$. Since \C$Q `<` P$, we can pick an element \C$e$ in \C${eq Q} `\` {eq P}$ and verify that the facet \C$F := 'P^=(base; [fset e])$ of \C$P$ satisfies \C$Q `<` F `<` P$. The general case where \C$Q'$ is a proper face of \C$P$ is handled by using the fact that \C$Q \in face_set P$ and \C$Q `<=` Q'$ ensures that \C$Q$ is a face of \C$Q'$ (see \C$Lemma |*face_set_of_face*|$).

\subsection{Vertices, Atomicity and Coatomicity}

The \emph{atoms} of a lattice $L$ are the elements $u \in L \setminus \{\bot\}$ such that there is no $v \in L$ satisfying $\bot \prec v \prec u$, where $\bot$ denotes the bottom element of $L$. In the face lattice of a polyhedron \C$P$, they correspond to the faces \C$F$ of \C$P$ such that \C$dim F = 1$, \ie, to the vertices of \C$P$ (remember
the shift by one of our formalization). This motivates the introduction of the vertex set of \C$P$, which satisfies the following two characteristic properties:
\begin{lstlisting}
Lemma |*in_vertex_setP*| (P : 'poly_n) (x : 'cV_n) :
  x \in vertex_set P <-> [pt x]%:PH \in face_set P.
Lemma |*face_dim1*| (P Q : 'poly_n) :
  Q \in face_set P -> dim Q = 1 ->
    exists2 x, Q = [pt x]%:PH & x \in vertex_set P.
\end{lstlisting}
where \C$[pt x]$ stands for the affine space reduced to the point \C$x$. A central property is that if \C$P$ is compact, then it coincides with the convex hull of its vertices:
\begin{lstlisting}
Theorem |*conv_vertex_set*| (P : 'poly_n) : compact P -> P = conv (vertex_set P).
\end{lstlisting}
Remark that this shows that any compact polyhedron is a polytope. Together with the converse statement (\C$Lemma |*compact_conv*|$ in \C$polyheron.v$), this brings a proof of Minkowski Theorem.

The latter result allows us to prove that, when \C$P$ is compact, its face lattice is atomistic, meaning that any face of \C$P$ is the join of a finite set of atoms:
\begin{lstlisting}
Lemma |*atomisticP*| (P : 'compact_n) (Q : face_set P) :
  reflect (exists2 S, (forall x, x \in S -> atom x) & Q = \join_(x in S) x)
          (atomistic Q).
Lemma |*face_atomistic*| (P : 'compact_n) (Q : face_set P) : atomistic Q.
\end{lstlisting}
where \C$'compact_n$ is the subtype of \C$'poly_n$ whose inhabitants are the compact polyhedra.
To prove this statement for \C$Q$, we define \C$S$ as the set of vertices of \C$Q$. They are vertices of \C$P$ as well, and thus correspond to atoms of the face lattice of \C$P$. We establish the inclusion \C$Q `<=` \join_(x in vertex_set Q) x$ by substituting \C$Q$ by \C$conv (vertex_set Q)$ thanks to \C$Theorem |*conv_vertex_set*|$ above, which makes the statement obvious by construction of the convex hull and the join operator. The proof is completed noting that the other inclusion (i.e., \C$Q `>=` \join_(x in vertex_set Q) x$) is trivial by the definition of the join operator.

The \emph{coatoms} of $L$ are defined dually: these are the elements $u \in L \setminus \{\top\}$ such that there is no $v \in L$ satisfying $u \prec v \prec \top$, where $\top$ denotes the top element of $L$. The coatomicity of \C$face_set P$ means that any face of \C$P$ is the intersection of facets of \C$P$. Our proof exploits the characterization of facets presented in Section~\ref{sec:gradedness}. We refer to \C$Lemma |*face_coatomistic*|$ for more details.

\subsection{Closedness under Taking Interval Sublattices}\label{subsec:interval}

The \emph{closedness under interval sublattices} of the face lattice of polytopes states that if \C$Q$ and \C$Q'$ are faces of a polytope \C$P$ such that \C$Q `<=` Q'$, then the \emph{interval} \C$'[< Q; Q' >]$, \ie, the sublattice of \C$face_set P$ composed of the faces \C$F$ satisfying \C$Q `<=` F `<=` Q'$, is isomorphic to the face lattice of a polytope of dimension \C$dim Q' - dim Q$. % chktex 13 chktex 8

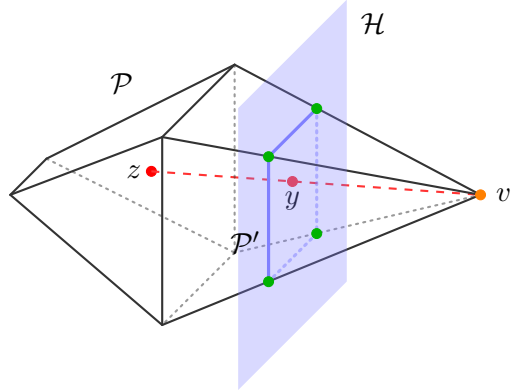
\begin{figure}[t]
\begin{center}
\begin{tikzpicture}[scale=1.25,line join=round,general/.style={fill=lightgray,draw=black,thin,fill opacity=.4},facet/.style={fill=blue!50},edge/.style={thick,fill opacity=0.4,line cap=round},vertex/.style={orange},back/.style={black!80,draw opacity=0.5,dotted},front/.style={black!80}]

\coordinate (v000) at  (0,0,0);
\coordinate (v001) at  (0,0,2);
\coordinate (v010) at  (0,2,0);
\coordinate (v011) at  (0,2,2);
\coordinate (v) at (3,1,1);
\coordinate (w) at (-2,1,0);
\coordinate (w') at (-2,1,1);

\path (v) node[right=2pt] {$v$};

\coordinate (x) at (-0.5,1.25,1);
\path (x) node[left] {$z$};
\filldraw[red] (x) circle (1.5pt);
\draw[red!80,thick,dashed] (x) -- (v);
\coordinate (y) at (1,1.142,1);
\path (y) node[below] {$y$};
\filldraw[red!80] (y) circle (1.5pt);

\draw[edge,back] (v000) -- (v001);
\draw[edge,front] (v001) -- (v011);
\draw[edge,front] (v011) -- (v010);
\draw[edge,back] (v010) -- (v000);
\draw[edge,back] (v) -- (v000);
\draw[edge,front] (v) -- (v001);
\draw[edge,front] (v) -- (v011);
\draw[edge,front] (v) -- (v010);
\draw[edge,back] (w) -- (v000);
\draw[edge,front] (w') -- (v001);
\draw[edge,front] (w') -- (v011);
\draw[edge,front] (w) -- (v010);
\draw[edge,front] (w) -- (w');
\path (-1.2,1.8) node {$\Pcal$};

\coordinate (p00) at (1,-0.5,-0.5);
\coordinate (p01) at (1,-0.5,2.5);
\coordinate (p10) at (1,2.5,-0.5);
\coordinate (p11) at (1,2.5,2.5);
\fill[blue!50,fill opacity=0.3] (p00) -- (p01) -- (p11) -- (p10) -- cycle;

\path (p10) node[below right=2pt] {$\Hcal$};

\coordinate (p00') at (1,0.33,0.33);
\coordinate (p01') at (1,0.33,1.66);
\coordinate (p10') at (1,1.66,0.33);
\coordinate (p11') at (1,1.66,1.66);
\draw[edge,back,very thick,blue!50] (p00') -- (p01');
\draw[edge,front,very thick,blue!50] (p01') -- (p11');
\draw[edge,front,very thick,blue!50] (p11') -- (p10');
\draw[edge,back,very thick,blue!50] (p10') -- (p00');
\path (1,1,2.3) node {$\Pcal'$};

\filldraw[vertex] (v) circle (1.5pt);
\filldraw[green!70!black] (p00') circle (1.5pt);
\filldraw[green!70!black] (p01') circle (1.5pt);
\filldraw[green!70!black] (p10') circle (1.5pt);
\filldraw[green!70!black] (p11') circle (1.5pt);
\end{tikzpicture}
\end{center}
\vskip-3ex
\caption{The vertex figure construction, illustrated on the vertex $v$ of the polytope $\Pcal$. The hyperplane $\Hcal$ (light blue) separates $v$ from the other vertices of the polytope. In the sliced polytope $\Pcal'$, the vertices (green) and edges (blue) are respectively in one-to-one correspondence with the edges and facets of $\Pcal$ containing $v$.}\label{fig:vertex_figure}
\end{figure}

The interest of this property is that it allows involved induction schemes on the height of the face lattice. As an example, we can establish the so-called \emph{diamond property}, namely that every interval of height $2$ of the face lattice consists of precisely four faces ordered as
\tikz[baseline=2mm,scale=0.3] {
\filldraw[black] (0,0) circle (5pt);
\filldraw[black] (-1,1) circle (5pt);
\filldraw[black] (1,1) circle (5pt);
\filldraw[black] (0,2) circle (5pt);
\draw (0,0) -- (-1,1) -- (0,2) (0,0) -- (1,1) -- (0,2); % chktex 8
}. Equivalently, this means that for any two faces $\Fcal$ and $\Fcal'$ of a polytope $\Pcal$ such that $\dim \Fcal' = \dim \Fcal + 2$ and $\Fcal \subset \Fcal'$, there are precisely two faces between them (see Figure~\ref{fig:face_lattice} for an illustration, and \C$Lemma |*diamond*|$ for the statement). The proof exploits the closedness under intervals, and the subsequent isomorphism of any interval of height $2$ with the face lattice of a polytope \C$P'$ verifying \C$dim P' = 2$. \C$Lemma |*dim2P*|$ of Section~\ref{sec:dim} reduces it to the face lattice of a segment \C$[segm x & y]$, which is given by the following characterization:
\begin{lstlisting}
Lemma |*face_set_segm*| (x y : 'cV_n) :
  face_set [segm x & y] = [fset [poly0]; [pt x]%:PH; [pt y]%:PH; [segm x & y]].
\end{lstlisting}

The proof of the closedness under interval sublattices is done as follows. First, we reduce to the case where \C$Q' = P$, since the face lattice of \C$Q'$ is isomorphic to the sublattice of \C$face_set P$ composed of the faces of \C$P$ contained in \C$Q'$. We are left with the following statement:
\begin{lstlisting}
Lemma |*closed_by_interval_r*| (P : 'compact_n) (Q : face_set P) :
  exists (P' : 'compact_n) (f : {omorphism '[< Q; P >] -> face_set P'}),
    bijective f & \pdim P = (\pdim P' + rank Q)%N.
\end{lstlisting}
where \C$rank$ is the rank function of the face lattice \C$face_set P$ which, as we already mentioned, is defined as the dimension of the face.
The proof is done by induction on the dimension of \C$Q$. We restrict the exposition to the base case \C$dim Q = 1$, \ie, when \C$Q$ corresponds to a vertex \C$v$ of \C$P$, since the general case is just handled by iterating the process. When \mbox{\C$dim Q = 1$}, the construction of the polyhedron \C$P'$ is achieved by the \emph{vertex figure} method. It consists in slicing the polytope \C$P$ with a hyperplane \C$[hp e]$ separating the vertex \C$v$ from the other vertices; see Figure~\ref{fig:vertex_figure} for an illustration. The hyperplane is defined by applying \C$Theorem separation$ (see Section~\ref{subsec:operations}) to the set \C$V$ consisting of the vertices of \C$P$ distinct from \C$v$. We define \C$P'$ as the sliced polytope. It has dimension \C$(dim P)-1$, and its face lattice can be shown to be isomorphic to the interval \C$'[< [pt v]%:PH; P >]$. Once again, the isomorphism is proved by exposing the polyhedron \C$P$ to the subtype \C${poly base}$ for some ambient representation \C$base$, and reducing to basic manipulations of sets \C${eq _}$ of active inequalities of faces. Interestingly, two distinct ambient representations are used in the proof: \C$base$ for the original polytope \C$P$, and its union \C$e +|` base$ with the singleton $\{\C$e$\}$ for the sliced polytope \C$P'$. Our use of canonical structures still applies to this setting, and provides the proof that any face of \C$P$ sliced with the hyperplane \C$[hp e]$ writes down over the base \C$e +|` base$ of the sliced polytope \C$P'$, as shown by the following typecheck:
\begin{lstlisting}
Variable (F : {poly base}).
Check ((slice e F)%:poly_base : {poly (e +|` base)}).
\end{lstlisting}

\section{A Formal Proof of Balinski's Theorem}\label{sec:balinski}
\tikzmark[baseline]{m7}Given a $n$-dimensional polytope $\Pcal$, Balinski's Theorem~\cite{Balinski61} states that the graph of $\Pcal$ is $n$-connected, which means that it is still a  connected graph after the removal of any $(n-1)$ vertices. In this section, we explain how the contributions presented in the previous sections allow us to formalize the proof of this result. Throughout this section, we fix a polyhedron \C$P : 'poly_n$ that satisfies \C$compact P$, \ie, \C$P$ represents a polytope.
\begin{tikzpicture}[remember picture,overlay,font=\scriptsize]%
\node[above left=0.7cm and 0.3cm, anchor=base east,text width=2cm,align=right] at (pic cs:m7) {\relevantfile{poly\_base.v}};%
\end{tikzpicture}%

We start by defining the adjacency relation between any two vertices \C$v$ and \C$w$ of \C$P$:
\begin{lstlisting}
Definition |*adj*| (P : 'poly_n) :=
	[rel v w : 'cV_n | (v != w) && ([segm v & w] \in face_set P)].
\end{lstlisting}
which corresponds to the fact that the segment between the two points \C$v$ and \C$w$ is an edge of \C$P$, \ie, a $1$-dimensional face. We do not need to require \C$v$ and \C$w$ to be vertices of \C$P$, since this is implied by the fact that \C$[segm v & w]$ is a face of \C$P$ with vertices \C$v$ and \C$w$:
\begin{lstlisting}
Lemma |*adj_vtxl*| (v w : 'cV_n) : adj P v w -> v \in vertex_set P.
Lemma |*adj_vtxr*| (v w : 'cV_n) : adj P v w -> w \in vertex_set P.
\end{lstlisting}

\subsection{Connectedness of the Adjacency Graph}

Before establishing that the adjacency graph is $n$-connected, we focus on the proof of its connectedness. This is carried out thanks to an abstract version of the simplex method which directly pivots over adjacent vertices (rather than over adjacent bases), up to finding a vertex of the optimal face associated with a certain cost function. In more detail, given a vector \C$c : 'cV_n$ representing a linear function \C$fun x => '[c,x]$, the pivoting rule is provided by the relation:
\begin{lstlisting}
Definition |*improve*| (v w : 'cV_n) := (adj P v w) && ('[c,v] > '[c,w]).
\end{lstlisting}
Thus, if the current vertex is \C$v$ and there exists \C$w$ such that \C$improve v w$ holds, \ie, such that \C$w$ is adjacent to \C$v$ and \C$'[c,w] < '[c,v]$, we pivot to \C$w$. If there is no such vertex, the following lemma ensures that \C$v$ minimizes the function \C$fun x => '[c,x]$:
\begin{lstlisting}
Lemma |*adj_argmin*| (v : 'cV_n) :
  v \in vertex_set P -> (forall w, adj P v w -> '[c,v] <= '[c,w]) ->
    v \in argmin P c.
\end{lstlisting}
We comment on the proof of \C$Lemma adj_argmin$, which relies on the vertex figure construction described in Section~\ref{subsec:interval}. Let $\Hcal$ be an hyperplane separating \C$v$ from the other vertices of \C$P$. Then, given an arbitrary point \C$z$ of \C$P$, the halfline $\ropen{\C$v$,\C$z$}$ intersects $\Hcal$ at a point \C$y$ (depicted in red in Figure~\ref{fig:vertex_figure}). As an element of $\Pcal \cap \Hcal$, \C$y$ belongs to the convex hull of the vertices of $\Pcal \cap \Hcal$ (depicted in green in Figure~\ref{fig:vertex_figure}). By the vertex figure construction, we know that the latter are precisely given by the intersection points of the edges incident to \C$v$ with the hyperplane $\Hcal$. Thanks to the assumption \C$(forall w, adj P v w -> '[c,v] <= '[c,w])$, the value of the function \C$fun x => '[c,x]$ on these vertices, and subsequently on \C$y$, is greater than or equal to \C$'[c,v]$. Since \C$z = v + t *: (y - v)$ for some \C$t >= 0$, we deduce that \C$'[c,v] <= '[c,z]$. % chktex 8

In this way, the abstract simplex method previously described allows us to build a path from any vertex \C$v$ of \C$P$ to the optimal face \C$argmin P c$. This is stated as:
\begin{lstlisting}
Lemma |*improve_path*| (v : 'cV_n) :
  v \in vertex_set P ->
    exists2 p, path improve v p & last v p \in vertex_set (argmin P c).
\end{lstlisting}
where \C$path improve v p$ means that \C$p$ is a sequence of points \C$w1$,\dots, \C$wq$ such that the properties \C$improve v w1$, \C$improve w1 w2$,\dots \ hold, and \C$last v p$ stands for the last element of the sequence \C$v :: p$.

In order to construct a path connecting any two vertices \C$v$ and \C$w$ of \C$P$, it now suffices to choose \C$c$ such that \C$w$ is the unique point of \C$P$ minimizing the linear function associated with \C$c$, \ie, such that \C$argmin P c = [pt w]%:PH$. This is possible since \C$w$ is a vertex, or equivalently, \C$[pt w]%:PH \in face_set P$; see \C$Lemma |*face_argmin*|$. We end up with the following statement:
\begin{lstlisting}
Theorem |*connected_graph*| (v w : 'cV_n) :
  v \in vertex_set P -> w \in vertex_set P ->
    exists2 p, path (adj P) v p & w = last v p.
\end{lstlisting}
which precisely corresponds to the connectedness of the graph of $\Pcal$.

\subsection{Generalization to the \texorpdfstring{$n$}{n}-Connectedness.}

On top of being compact, we suppose that $\Pcal$ has dimension $n$, \ie, \C$dim P = n.+1$. In this case, the $n$-connectedness of the graph of $\Pcal$ writes as follows:
\begin{lstlisting}
Theorem |*balinski*| (V : {fset 'cV_n}) (v w : 'cV_n) :
  (V `<=` vertex_set P)%fset -> #|` V | = n.-1 ->
    v \in (vertex_set P `\` V)%fset -> w \in (vertex_set P `\` V)%fset ->
      exists p, [/\ (path (adj P) v p), w = last v p & all [predC V] p].
\end{lstlisting}
Here, the finite set \C$V$ stands for the subset of the vertices of \C$P$ of cardinality $n-1$ that we remove from the graph, and \C$v$ and \C$w$ are any two vertices not belonging to \C$V$. The property \C$all [predC V] p$ of the path \C$p$ from \C$v$ to \C$w$ corresponds to the fact that no vertex of \C$p$ belongs to the set \C$V$.

\begin{figure}
\begin{center}
\tdplotsetmaincoords{70}{110}
\begin{tikzpicture}[tdplot_main_coords,line join=round,edge/.style={thick,fill opacity=0.4,line cap=round},back/.style={black!80,draw opacity=0.5,dotted},front/.style={black!80},>=stealth',decoration={
markings,mark=at position 0.5 with {\arrow{>}}}]
\coordinate (b1) at (-1.5,-1,0);
\coordinate (b2) at (1,-1.3,0);
\coordinate (b3) at (1.2,1,0);
\coordinate (b4) at (-1.1,1.5,0);

\coordinate (t1) at (-0.7,-1,4);
\coordinate (t2) at (0.8,-0.9,4);
\coordinate (t3) at (1.2,1.1,4);
\coordinate (t4) at (-1.2,1.2,4.5);

\coordinate (m1) at (-2.5,-2,1.8);
\coordinate (m2) at (2,-2,2.5);
\coordinate (m23) at (3,-0.5,2.3);
\coordinate (m3) at (2.1,2.4,2);
\coordinate (m4) at (-2,1.8,1.5);

\coordinate (h1) at (-1.2,-1.7,4);
\coordinate (h3) at (2,2,4);
\coordinate (h5) at (-1.333,1.3,4);

\coordinate (d1) at ($(h5)-(t1)$);
\coordinate (d2) at ($(h5)-(t3)$);
\path[name path=s14] (h1) -- ($(h1) + 2*(d1)$);
\path[name path=s34] (h3) -- ($(h3) + 2*(d2)$);
\path[name path=s12] (h1) -- ($(h1) - 2*(d2)$);
\path[name path=s32] (h3) -- ($(h3) - 2*(d1)$);
\path[name intersections={of=s14 and s34, by=i}] coordinate (h4) at (i);
\path[name intersections={of=s12 and s32, by=i}] coordinate (h2) at (i);

\fill[green!70!black,fill opacity=.3] (b1) -- (b2) -- (b3) -- (b4) -- cycle;
\draw[edge,back] (b4) -- (b1) -- (b2);
\draw[edge,front] (b2) -- (b3) -- (b4);
\draw[edge,front] (t1) -- (t2) -- (t3) -- (t4) -- cycle;
\draw[edge,front] (m2) -- (m23) -- (m3) -- (m4);
\draw[edge,back] (m4) -- (m1) -- (m2);

\draw[edge,front] (m2) -- (t2);
\draw[edge,back] (t1) -- (m1);
\draw[edge,front] (m3) -- (m4) -- (t4) -- (t3) -- cycle;

\draw[edge,front] (b2) -- (m2);
\draw[edge,back] (b1) -- (m1);
\draw[edge,front] (b3) -- (b4) -- (m4) -- (m3) -- cycle;
\draw[edge,front] (m23) -- (t2) -- (t3) -- cycle;
\draw[edge,front] (m23) -- (b2) -- (b3) -- cycle;

\draw[orange,very thick,postaction={decorate}] (t2) -- (m2);
\fill[blue!50,fill opacity=0.32] (h1) -- (h2) -- (h3) -- (h4) -- cycle;
\draw[edge,front] (t1) -- (t2) -- (t3) -- (t4) -- cycle;
\draw[edge,front] (h5) -- (t4);
\draw[blue!50,thick,dashed] (t1) -- (t2) -- (t3) -- (h5) -- cycle;
\draw[green!70!black,very thick,postaction={decorate}] (b2) -- (b3);
\draw[green!70!black,very thick,postaction={decorate}] (b3) -- (b4);
\draw[orange,very thick,postaction={decorate}] (m2) -- (b2);
\draw[red!80,very thick,postaction={decorate}] (m4) -- (b4);

\filldraw[orange] (t2) circle (2pt) node[left,black] {$v$};
\filldraw[green!70!black] (b2) circle (2pt) node[left,black] {$x$};
\filldraw[green!70!black] (b4) circle (2pt) node[right,black] {$y$};
\filldraw[red!80] (m4) circle (2pt) node[right,black] {$w$};
\filldraw[blue!50] (t3) circle (2pt) node[right,black] {$v^1$};
\filldraw[blue!50] (t1) circle (2pt) node[above left=-1ex and .25ex,black] {$v^2$};
\path (h4) ;
\draw[edge,very thick,<-] (0,-4,2.5) -- (0,-4,1.5);
\path (0,-4,2.5) node[left=.7ex,black] {$c$};
\end{tikzpicture}
\end{center}
\caption{Illustration of the proof of Balinski's Theorem, where the set $V$ of removed vertices is $\{v^1, v^2\}$. The hyperplane satisfying   Conditions~\ref{item:conf1} and~\ref{item:conf2} is depicted in light blue, the optimal face \C$F = argmin P c$ in green, and the three paths \C$p1$, \C$p2$ and \C$p3$ respectively in orange, red and green.}\label{fig:balinski}
\end{figure}
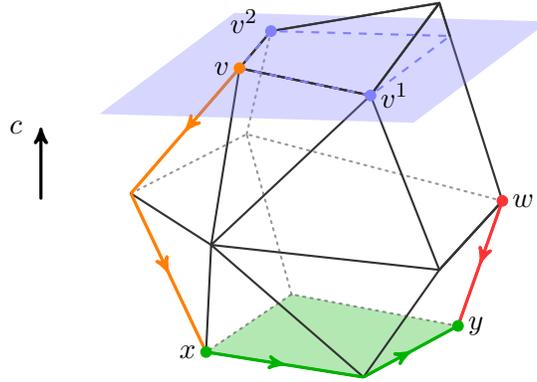

The proof relies on the existence of a hyperplane \C$[hp (c,alpha)]$ such that:
\begin{enumerate}[(i)]
\item\label{item:conf1} the vertex \C$v$ and the set \C$V$ are both contained in \C$[hp (c,alpha)]$;
\item\label{item:conf2} the vertex \C$w$ satisfies \C$'[c,w] <= alpha$, and there exists a point \C$z$ of \C$P$ such that \C$'[c,z] < alpha$.
\end{enumerate}
An example of such a configuration is depicted in Figure~\ref{fig:balinski}, taking \C$z$ as the vertex \C$w$. We introduce the optimal face \C$F = argmin P c$ associated with the cost vector \C$c$. It is immediate to show that for any \C$x \in F$, we have \C$'[c,x] < alpha$, since \C$'[c,x] <= '[c,z] < alpha$. We build a path from \C$v$ to \C$w$ in \C$(vertex_set P) `\` V$ by concatenating three paths:
\begin{itemize}
\item two paths \C$p1$ and \C$p2$, provided by \C$Lemma improve_path$, respectively connecting \C$v$ and \C$w$ to some vertex of the optimal face \C$argmin P c$, say \C$x$ and \C$y$;
\item a path \C$p3$ from \C$x$ to \C$y$ contained in the graph of \C$argmin P c$, which is built by applying \C$Theorem connected_graph$ to \C$argmin P c$ instead of \C$P$.
\end{itemize}
We refer to Figure~\ref{fig:balinski} for an illustration. We claim that none of these paths intersects the set \C$V$. Indeed, the path \C$p1$ satisfies \C$path (improve v p1)$, hence \C$'[c,u] < '[c,v] <= alpha$ for any vertex \C$u$ in \C$p1$. The same argument applies to the path \C$p2$ (replacing \C$v$ by \C$w$). Finally, the face \C$F$ cannot intersect \C$V$, as \C$'[c,u] < alpha$ for all \C$u \in F$. Moreover, the graph of \C$F$ is included in the graph of \C$P$; see \C$Lemma |*sub_adj*|$ that is a consequence of the fact that the faces of \C$F$ are faces of \C$P$. We deduce that \C$p3$ is a path in the graph of \C$P$ with empty intersection with \C$V$.
As a result, the concatenation \C$(p1 ++ p3 ++ (rev (belast w p2)))$ provides the expected path from \C$v$ to \C$w$ (the term \C$belast w p2$ stands for the sequence \C$w :: p2$ minus the last element of \C$p2$, and \C$rev _$ for its reversal).

It remains to explain how we can ensure that Conditions~\ref{item:conf1} and~\ref{item:conf2} are met. The hyperplane \C$[hp (c,alpha)]$ is provided by the following lemma applied to \C$W = v |` V$.
\begin{lstlisting}
Lemma |*subset_hp*| (W : {fset 'cV_n}) :
  (0 < #|` W | <= n) -> exists2 e : lrel_n, (e.1 != 0) & {subset W <= [hp e]}.
\end{lstlisting}
This lemma states that we can always find a (proper) hyperplane containing any nonempty set of at most $n$ points. Defining \C$c := e.1$ and \C$alpha := e.2$, the hyperplane \C$[hp e]$ consists of the points \C$x$ such that \C$'[c,x] = alpha$. Thus, Condition~\ref{item:conf1} is satisfied. In the case where \C$'[c,w] != alpha$, we arrive at \C$'[c,w] < alpha$ up to changing \C$e$ into \C$-e$ (\ie, \C$c$ into \C$-c$ and \C$alpha$ into \C$-alpha$, which does not affect \C$[hp e]$. If \C$'[c,w] = alpha$, we exploit the fact that the polyhedron \C$P$ cannot be contained in \C$[hp e]$, since \C$dim P = n.+1$ while \C$dim [hp e] = n$ (see \C$Lemma dim_hp$). Thus, there exists a point \C$z$ of \C$P$ such that \C$'[c,z] != alpha$. Up to changing \C$e$ into \C$-e$, this ensures that Condition~\ref{item:conf2} is satisfied.

\section{Related Work}\label{sec:related_work}

Many software developments related with convex polyhedra have been motivated by applications to formal verification. Several libraries have been developed for this purpose, \eg{}~\cite{PPL,Apron}, and, despite being informal, it is worth noting that they are also used by mathematicians to perform computation over polyhedra and polytopes, for instance in~\cite{polymake:2000,sage}. Initiatives on the development of formally verified polyhedral algorithms are more recent. The works of~\cite{Spasic2012} and~\cite{BotteschFROCOS2019} in Isabelle/HOL aim at providing a formally proven yet practical and efficient algorithm to decide linear rational arithmetic for SMT-solving. The Micromega tactics~\cite{Besson2007} relies on polyhedra to prove automatically arithmetic goals over ordered rings in~\Coq{}. The \emph{Verified Polyhedral Library}~\cite{Boulme2018,Fouilhe2014} targets abstract interpretation, and brings the ability to verify polyhedral computations a posteriori in \Coq{}.

There are far fewer developments focusing on formal mathematics. Euler formula, which relates the number of vertices, edges and facets of three-dimensional polytopes, has been proved in~\cite{DufourdTCS2008} in~\Coq{} and in~\cite{AlamaICMS2010} in Mizar. The generalization to polytopes in arbitrary dimension, namely Euler--Poincaré formula, has been formally proved in HOL-Light~\cite{Harrison2013}, together with several intermediate properties of polyhedra and faces. In the intuitionistic setting, we are not aware of any work concerning faces and their properties. We point out that Fourier--Motzkin elimination has been formalized in~\Coq{} by~\cite{Sakaguchi2016}. % chktex 8

\section{Conclusion and Future Work}\label{sec:conclusion}

In this work, we have formalized a substantial part of the theory of polyhedra and their faces, which has allowed us to obtain some of the essential properties of face lattices. Beyond the mathematical results formally proven, a special attention has been paid to the usability of the library. This goes through a mechanism to bring the right representation of faces according to the context, and the automatic proof that these representations are valid thanks to the use of canonical structures.

This foundational work opens several perspectives. First, it has raised that an important development over ordered structures is still needed, in particular for the manipulation of ordered substructures such as sublattices, and the interplay between them through morphisms. The formalization of finite groups and subgroups in~\cite{GonthierITP2013} may provide a possible source of inspiration to solve this problem. Second, there are many other interesting properties in relation with polyhedra and their faces to be formalized, such as getting upper bounds on the diameter of polytopes, or more generally, on the number of faces (the so-called f-vector theory). However, beyond the interest of formalizing already known mathematical results, we are even more interested in using proof assistants to help getting new ones. We think of mathematical results relying on computations that are not accessible by hand. To this extent, we aim at providing a way to compute with the objects introduced in this work, directly within the proof assistant, and to introduce all the needed mechanisms for the design and development of large scale reflection tactics. A basic goal is to compute the face lattice (or part of it) of a polyhedron defined by a set of inequalities. This requires us to formalize some algorithms based on faces, and to find a way to execute them on efficient data structures, in the spirit of the approach of~\cite{Cohen2013}.

\section*{Acknowledgment}
\noindent We are grateful to Assia Mahboubi for helpful discussions on the subject. We warmly thank Kazuhiko Sakaguchi for his help on integrating the ordered structures of \MathCompShort{}, and Quentin Canu for discussions on the formalization of affine hulls. We finally thank the  reviewers for their comments and suggestions  (including those of the IJCAR 2020 conference paper).

  %% the following bibliography is gererated manually for the sake of brevity
  %% only; please use a separate .bib file in your submission

\bibliographystyle{alphaurl}
\bibliography{journal}

\end{document}